\documentclass[showpacs,aps,pra,twocolumn]{revtex4}
\usepackage{amsmath}
\usepackage{amssymb}
\usepackage{graphicx}
\usepackage{ulem}
\usepackage{color}
\usepackage[dvipsnames]{xcolor}
\definecolor{orange}{rgb}{1,0.5,0}

\begin{document}

\title{Avoiding dissipation in a system of three quantum harmonic oscillators}
\author{Gonzalo Manzano$^{1,2}$}
\author{Fernando Galve$^1$}
\author{Roberta Zambrini$^1$}
\affiliation{$^1$ IFISC (UIB-CSIC), Instituto de F\'isica Interdisciplinar y Sistemas Complejos, UIB Campus, E-07122 Palma de Mallorca, Spain}
\affiliation{$^2$ Departamento de F\'isica At\'omica, Molecular y Nuclear, Universidad Complutense de Madrid, 28040-Madrid, Spain} 
\date{\today}

\begin{abstract}
We analyze the symmetries in an open quantum system composed by three coupled and detuned harmonic oscillators in the presence of a common 
heat bath. It is shown analytically how to engineer the couplings and frequencies of the system so as to have several degrees of freedom
unaffected by decoherence, irrespective of the specific spectral density or initial state of the bath. This partial thermalization allows observing asymptotic 
entanglement at moderate temperatures, even in the nonresonant case. This latter feature cannot be seen in the simpler situation of only two oscillators, 
highlighting the richer structural variety of the three-body case. When departing from the strict conditions for partial thermalization, a hierarchical structure 
of dissipation rates for the normal modes is observed, leading to a long transient where quantum correlations such as the quantum discord are largely preserved, 
as well as to synchronous dynamics of the oscillators quadratures.
\end{abstract}

\maketitle

\section{Introduction}

 Prevention of decoherence and dissipation in open quantum systems is a fundamental condition for the presence of
quantum phenomena in warm macroscopic every-day world. While decoherence has been extensively studied from the early 
1980s \cite{Zurek1,Zurek2} to nowadays (for a list of reviews see for instance
\cite{decoherence1,decoherence2,decoherence3,decoherence4}), and different mechanisms to avoid it have been discussed
in the last years, including strategies to engineer it for applications \cite {Diehl,cirac2009,blatt2011}. Furthermore, some 
macroscopic systems from  marine algae \cite{algae} to metal carboxylates \cite{carboxylates}, can present quantum correlations at 
high temperatures. These recent  experiments suggest that avoiding a complete quantum-to-classical transition induced by monitoring of 
the environment 
is not only artificially possible but can occur inherently in natural phenomena. The mechanisms that produce such survival or even 
construction of coherence at large time scales remain almost unclear, but different theoretical strategies have been proposed in
order to predict it, mostly  motivated in the context of quantum computation \cite{strategies,strategies2,galvePRL,DFSs,IPSs}. 
Indeed decoherence and the leak of information to the environment are the major obstacle in quantum processing of 
information and construction of quantum memories \cite{Chuang}.

In this context, one of the strategies to bypass decoherence is exploiting dynamical symmetries in the
system-environment interaction. In  order to generate unitary evolution in a certain subspace of the Hilbert
space of an open system, a common dissipation (where  several units equally couple to the same environment)
has been firstly used in a two-qubit system \cite{Zanardi1,Duan,ent,ent2} and later extended to multiple 
qubits \cite{3spins,3spins2} and continuous variable systems \cite{first,Liu,Chu,Paz,Galve,Ficek,moussa,our}. A
general framework has been developed with several contributions (see for example \cite{DFSs} and the references 
therein) agglutinating the main concepts of decoherence-free subspaces (DFSs), Noiseless subsystems (NSs) or more 
recently information-preserving structures (IPSs) \cite{IPSs} and proposing general methods to obtain them for 
arbitrary Hamiltonians. In our work we extend previous studies in the context of continuous variables exploring 
the vaster landscape offered when moving from two to three coupled harmonic oscillators in the search for NSs induced by non-trivial 
symmetries in the system. 

Previous works on dissipative harmonic oscillators reported that in presence of identical frequencies 
and couplings between oscillators the symmetry of the collective motion and their interaction with
the environment can lead to the effective decoupling of some of the normal modes of the system from the 
bath \cite{first,Liu,Chu,Paz,Ficek}. The cases of different frequencies \cite{Galve,alonso} 
or couplings \cite{moussa} is 
instead less studied and  understood. The natural step of considering three harmonic oscillators beyond the 
symmetric configuration of identical oscillators already provides much more phenomenological richness 
while at the same time it allows for analytic treatment and gives valuable intuition when pursuing a further 
extension to the case of $N$ oscillators. 

We show in this work how to obtain NSs and bypass decoherence independently of the bath properties such  as
temperature or frequency distribution, considering different frequencies, couplings and boundary conditions,
in the presence  of a common bath. We also analyze how by using these NSs, quantum correlations like
entanglement can persist (by two different mechanisms) in the asymptotic limit of the dynamical evolution,
given in our case by a Markovian-type dynamics in the weak-coupling limit. Furthermore even for three
different oscillators an homogeneous dynamics, such as collective quantum synchronization, is shown to be
present in some parameters manifolds. Synchronization is a widespread phenomenon in physical, biological, chemical
and social systems \cite{pik}, only recently  explored in the quantum regime \cite{our,syncnet}. Our results for the
system of three oscillators in presence of common dissipation may be implemented with ions in linear Paul
traps by following the proposal in \cite{Serafini}.  Experimental realization of coupled harmonic oscillators
appear in optical \cite{opt,opt2} and superconducting  \cite{superc,superc2} cavities as well as trapped ions
\cite{ions,ions2} or nanoelectromechanical resonators \cite{Plenio}.  Three coupled elements architectures are
also known to allow for isochronous synchronization of semiconductor lasers with delayed  coupling  or
neuronal  models \cite{Fischer}.

The paper is organized as follows: in Sect. \ref{model} we introduce the model for the system of three harmonic
oscillators dissipating  into a common bath, also in terms of the normal modes of the system.  For certain particular
values of the system's parameters (and independently of the  bath characteristics), one of several normal modes can be
protected from decoherence. We find analytically these conditions leading for NSs in Sect. \ref{decoherence-free} giving 
some specific cases that are analyzed in more detail. In the following we focus on the long time dynamics of the system
considering in Sect. \ref{asymptotic_ent} the creation and conservation of entanglement at asymptotic times. Section \ref{sync} 
is dedicated to ability of the system to generate asymptotic synchronized states and exploring its connection to the 
robustness of quantum correlations. In section \ref{robustness} deviations of the NSs conditions are considered. This leads to dynamical relaxation 
of the system that converges towards a thermal state. We conclude with Sect. \ref{conclusions} summarizing our main results.

\section{The model} \label{model}
We start with a Hamiltonian describing three coupled quantum harmonic oscillators with arbitrary frequencies and coupling constants between 
them. For simplicity we suppose unit masses and natural units $(\hbar = k_B = 1)$:
\begin{equation}
 H_S = \frac{1}{2} \sum_{i=1}^N \left( p_i^2 + \omega_i^2 q_i^2 \right) + \sum_{i < j} \lambda_{i j} q_i q_j
\end{equation}
 where $p_i$ and $q_i$ represents respectively the momentum and position operators of each harmonic oscillator $([q_i,p_j]= i \delta_{i j})$ 
 and $N=3$ for our case. This equation is conveniently expressed in quadratic matrix form as:
\begin{equation}\label{H_S}
 H_S= \frac{1}{2} \left( {\bf{p}}^T \mathbb{I}  {\bf{p}}  + {\bf{q}}^T \mathcal{H}  {\bf{q}} \right)
\end{equation}
where $\mathbb{I}$ is the identity $(N \times N)$ matrix, ${\bf{q}}^T = (q_1, ..., q_N)$ and $\mathcal{H}$ contains all the parameters of the 
system, i.e. the squared frequencies and couplings between oscillators. We will only consider $\mathcal{H}$ with positive eigenvalues, 
so as to have bounded states (attractive potential). 

The environment is introduced by equally coupling each oscillator of the system to the same thermal 
bath, which is described by an infinite collection of independent bosonic modes:
\begin{equation} 
 H_B = \frac{1}{2} \sum_{\alpha=1}^\infty \left( \Pi_\alpha^2 + \varOmega_\alpha^2 X_\alpha^2 \right) 
\end{equation}
where  $[X_\alpha , \Pi_\beta] = i \delta_{\alpha \beta}$ and masses are unitaries. We will use along the paper 
greek subindices to refer to bath modes while latin ones are reserved for system oscillators ($i, j$) and normal modes ($k, n$). The 
system-bath interaction reads:
\begin{equation}\label{H_I}
 H_I = \sum_{i=1}^N q_i \sum_{\alpha=1}^\infty \lambda_\alpha X_\alpha,
\end{equation}
with a factorized form $H_I = S \otimes B$ of an operator $S$ acting only on the system's Hilbert space and $B$ acting on
the environment one. As usual, this type of interaction yields a renormalization of the frequencies that we may include
directly in our model  by changing $\omega_i^2 \rightarrow \omega_i^2 + \sum_\alpha \frac{\lambda_\alpha^2}{2
\varOmega_\alpha^2}$ \cite{Breuer}.

The normal modes basis of the system (\ref{H_S})
is obtained after a canonical transformation of the system operators through the orthogonal
basis-change matrix $\mathcal{F}$:
\begin{equation}\label{basis_change}
 q_i = \sum_{k=1}^N \mathcal{F}_{i k} Q_k ~~~;~~~ p_i = \sum_{k=1}^N \mathcal{F}_{i k} P_k
\end{equation}
which diagonalizes $\mathcal{H}$ $({\bf q}^T \mathcal{H} {\bf q} = {\bf Q}^T {\Omega} {\bf Q})$. Here 
$\Omega= \mathcal{F}^T \mathcal{H} \mathcal{F}$ is a diagonal matrix containing the squared normal 
modes frequencies $\Omega_n$ with $n=1,2,..., N$. In this basis $H_S$ now represents the Hamiltonian for a $N=3$ uncoupled harmonic 
oscillators, or normal modes, related with the original (natural) modes by $\mathcal{F}$, so we can write:
\begin{equation}
 H_S = \frac{1}{2} \sum_{n=1}^N \left( P_n^2 + \Omega_n^2 ~Q_n^2 \right)
\end{equation}
 and the interaction Hamiltonian of Eq. (\ref{H_I}) is now:
\begin{equation}\label{interaction}
  H_I = \sum_{n=1}^N \kappa_n Q_n \sum_{\alpha=1}^{\infty} \lambda_{\alpha} X_{\alpha}.
\end{equation}
We stress that even if the oscillators are coupled with the same strength to the bath, 
the couplings of the normal modes to the bath, $\kappa_n$, are not homogeneous:
\begin{equation}\label{kappas}
 \kappa_n = \sum_{i=1}^N \mathcal{F}_{i n}.
\end{equation}
The {\it effective couplings} $\kappa_n$  only depend on the canonical transformation, i.e. on the system's 
parameters and arrangement defined by $\mathcal{H}$. This suggests
 a strategy to protect one or more normal modes from the environment action based on proper tuning of 
 $\{\omega_i,\lambda_{ij}\}$. Our analysis in section \ref{decoherence-free} addresses this point 
while deviations form the condition of vanishing effective coupling of a system normal mode and the environment
are explored in section \ref{robustness}.

We mention that while this work focuses on the case of three coupled harmonic oscillators, the description in
terms of effective  couplings here provided is rather general and applies for arbitrary networks of $N$ harmonic
oscillators. The case of a common bath  for all oscillators in the system corresponds to situations where the
correlation length in the environment is larger than the system size.  This assumption is not crucial for the
framework we develop here, though any other choice would produce different specific analytic expressions. Finally, the case
of a separate bath for each oscillator was shown in Ref. \cite{our} to yield  equal decoherence for all normal
modes and therefore no NSs nor synchronization.

\subsection{Unbalanced bath couplings}
While the common bath situation is mostly a characteristic related to the coherence length of the bath as compared 
to the extension of the system, the equal coupling of each system oscillator to the latter might seem an arbitrary restriction.
Imagine for example that each oscillator is at a different distance from the common heat bath, leading to an interaction
\begin{equation}
H_I=\sum_{i=1}^N \gamma_i q_i \sum_{\alpha=1}^\infty\lambda_\alpha X_\alpha
\end{equation}
where the different oscillators feel a coupling of strength $\gamma_i$. The immediate consequence is that the effective couplings become
\begin{equation}
\kappa_n = \sum_{i=1}^N \gamma_i \mathcal{F}_{i n}.
\end{equation}
Though in this paper we will consider $\gamma_i=1$, the unbalanced case would be solved following exactly the same procedure as we outline in the next
section.

\section{Noiseless subsystems and asymptotic properties}

\label{decoherence-free}

In this section we discuss the conditions to achieve noiseless subsystems where 
dissipation is avoided in one or two of the system's normal modes. The properties of our system are specified completely by the matrix 
$\mathcal{H}$ appearing in Eq. (\ref{H_S}):
\begin{equation}
 \mathcal{H} = \left( \begin{array}{cc}
                        \omega_1^2 ~~ \lambda_{1 2} & \lambda_{1 3} \\
			\lambda_{1 2} ~~ \omega_2^2  & \lambda_{2 3} \\
			\lambda_{1 3} ~~ \lambda_{2 3}  & \omega_3^2 \\
                       \end{array} \right)
\end{equation}
and we aim to derive the set
of conditions for the system parameters leading to one or two normal modes decoupled from the environment, i.e.
whose effective coupling $\kappa_n$ is zero. 

Let us consider a normal mode $\delta$ with normal frequency $\Omega_\delta$. The eigenvalue problem is
expressed adequately by three equations:
\begin{equation}\label{eqs3}
 (\mathcal{H} - \Omega_\delta^2 \mathbb{I}){\bf C}_{\delta} = 0
\end{equation}
for the components of the normal vector 
${\bf C}_\delta=( \mathcal{F}_{1 \delta},\mathcal{F}_{2 \delta}, \mathcal{F}_{3 \delta})^T$ with  $\mathcal{F}_{ij}$
defined in Eqs.(\ref{basis_change}). The condition for normal mode $\delta$ to be non-dissipative (out of the bath' influence) 
leads to a constraint as follows
\begin{equation}\label{non-dissipative}
 \kappa_\delta=0 ~\Leftrightarrow~ \mathcal{F}_{1 \delta} + \mathcal{F}_{2 \delta} + \mathcal{F}_{3 \delta} = 0.
\end{equation}

From Eqs. (\ref{eqs3}), (\ref{non-dissipative}) and normalization condition we can obtain analytically ${\bf C}_\delta$, $\Omega_\delta$ 
with a further constraint for the system parameters. In other words, not all parameter choices lead to NSs, 
but it is possible for some configurations of frequencies and couplings of the set of three oscillators (satisfying some constraint).

The normal mode $\delta$ in terms of the system parameters reads
\begin{equation}
  {\bf C}_\delta = c \left( \begin{array}{cc}
                        \lambda_{1 3} \lambda_{1 2} + \lambda_{2 3} (\Omega_\delta^2 - \omega_1^2)\\
			(\Omega_\delta^2 - \omega_2^2)(\Omega_\delta^2 - \omega_1^2) - \lambda_{1 2}^2\\
			 \lambda_{1 3} \lambda_{2 3} + \lambda_{1 2} (\Omega_\delta^2 - \omega_3^2) \\
                       \end{array} \right)
\end{equation}
where $c$ is the  normalization constant. Applying Eq. (\ref{non-dissipative}) we can obtain its eigenfrequency $\Omega_\delta^2$ as:
\begin{eqnarray} \label{normal_freq}
 \Omega_\delta^2 = & \left( \frac{\omega_1^2 + \omega_3^2}{2} \right) - \left( \frac{\lambda_{1 2} + \lambda_{2 3}}{2}
  \right)  \pm    
 \sqrt{\Delta^2 + \left( \frac{\lambda_{1 2} + \lambda_{2 3}}{2}\right)^2}&  \\ \nonumber 
 &\overline{+ \Delta(\lambda_{2 3} - \lambda_{1 2}) + \lambda_{1 3}(\lambda_{1 3} - \lambda_{1 2} - \lambda_{2 3}) }&
\end{eqnarray}
where $\Delta =(\omega_1^2 - \omega_3^2)/2$. Finally, by defining the quantities $\Sigma = (\omega_1^2 + \omega_3^2)/2 - \omega_2^2$ 
and $\mathcal{R} = - \left( \lambda_{1 2} + \lambda_{2 3}\right)/2  \pm  \sqrt{\left( \Delta + (\lambda_{1 2} + 
\lambda_{2 3})/2 - \lambda_{1 3} \right)^2 + 2 \Delta(\lambda_{1 3} - \lambda_{1 2})}$ the constraint relation ($\kappa_\delta=0$) reads:
\begin{eqnarray} \label{ligadure}
 2 \lambda_{1 2} \lambda_{2 3} \mathcal{R} + \lambda_{1 3} (\lambda_{1 2}^2 + \lambda_{2 3}^2) +\lambda_{1 3}^2 ( \mathcal{R} + 
\Sigma)  &+& \nonumber \\
 - (\mathcal{R} + \Delta)( \mathcal{R} - \Delta )( \mathcal{R} + \Sigma) &= &0
\end{eqnarray}
Eq. (\ref{ligadure}) is one of our main results, and represents a hypersurface in the $d$-dimensional parameters space 
(being  $d=(N + 1)N/2 = 6$) whereby a normal mode is allowed to evolve freely and without dissipation. 
Such manifold is restricted to regions in which the normal mode frequency $\Omega_\delta$ is real and positive,
and the normalization constant $c$ is  well defined. It's worth noting that looking only to non-dissipative modes
(imposing the condition $\kappa_\delta = 0$) has been crucial in  order to solve analytically the above equations,
otherwise we have to deal with complicated expressions involving third order equations  (general normal mode
expression).

When Eq. (\ref{ligadure}) is fulfilled we obtain a NS composed by (at least) a single normal mode that is effectively
uncoupled  to the reservoir. This could be performed artificially by tuning one of the $d=6$ parameters of
$\mathcal{H}$, such as, for instance, the natural frequency of one oscillator. In experiments where local addressing is possible, such
as ions confined to individual traps, this modification  should be rather straightforward (see e.g. \cite{wineland2012}). It should be stressed
that noise models for ion traps typically favor a SB interpretation in terms of fluctuating uncorrelated surface dipoles \cite{dipoles}, though other 
microscopic models based on charge diffusion \cite{hankel} in the electrode surface question whether the bath's correlation length could in fact
be larger than the distance of the ion to the electrode. For the moment, this is an open problem.

Configurations in which a NS consisting of two normal modes is produced can  also be
obtained analytically 
in the specific case of three oscillators. Indeed, when two normal modes uncouple from the environmental action, the third one must necessarily coincide with the 
center of mass of the system. Explicitly, the condition for the center of mass being a normal mode is:
\begin{equation}\label{eq}
 (\mathcal{H} - \Omega_{CM}^2 \mathbb{I}){\bf C}_{CM} = 0  ~\Longleftrightarrow~ \Omega_{CM}^2 = \omega_i^2 + \sum_{j \neq i} \lambda_{i j} ~~\forall i
\end{equation}
with ${\bf C}_{CM} = (1, 1, 1)^T/\sqrt{3}$. The latter implication can be captured in the next two relations that 
have to be fulfilled simultaneously by the system parameters of the system:
\begin{eqnarray}\label{2modesA}
 \omega_1^2 = \omega_2^2 + \lambda_{2 3} - \lambda_{1 3} \\
\label{2modesB} \omega_3^2 = \omega_2^2 + \lambda_{1 2} - \lambda_{1 3}
\end{eqnarray}
Since we want to remain in the domain of attractive potentials, we have to restrict ourselves to regions of the parameter space where  
$\Omega_{CM}^2 = \omega_1^2 + \omega_3^2 - \omega_2^2 + 2\lambda_{1 3}>0$.

\begin{figure}[t]
 \includegraphics[width=8cm]{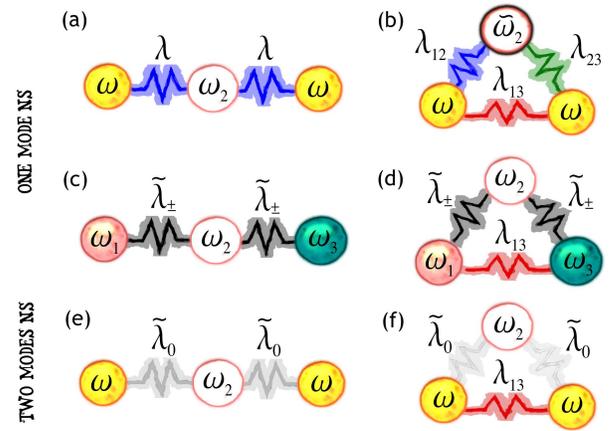}
 \caption{(Color online) Different configurations for a chain of three coupled oscillators
  in which a NS of one (a-d) or two (e-f) normal modes is predicted. 
  The tilde on parameters indicates a  fixed value depending 
  on the other non-tilded ones as described in the text. \label{figure1}}
\end{figure}

In order to see the power of the conditions
(\ref{ligadure}) and (\ref{2modesA},\ref{2modesB}) we will give in the following some
examples of  configurations in which a NS of one or two modes is produced. We will consider for instance simpler
situations in which the six-dimensional  parameter space is reduced by assuming two of the three natural frequencies of
oscillators to be equal, $(\omega \equiv \omega_1 = \omega_3  \neq \omega_2)$ or in which two of the three couplings are
equal $(\lambda \equiv \lambda_{1 2} = \lambda_{2 3} \neq \lambda_{1 3})$. This is  sufficient to obtain some different
scenarios appearing in open an closed chains configurations, as is schematically shown in Figure  \ref{figure1}.

Let's start from the case of two equal frequencies $(\omega \equiv \omega_1 = \omega_3 \neq \omega_2)$. Then the quantities defined in 
Eq. (\ref{ligadure}) are simply $\Delta = 0$, $\Sigma = \omega^2 - \omega_2^2$ and $\mathcal{R} = \{ - \lambda_{1 3} , \lambda_{1 3} - 
\lambda_{ 1 2} - \lambda_{2 3} \}$, the latter implying  two different consistent solutions to Eq. (\ref{ligadure}): the first one $ \{ 
\lambda_{1 2} = \lambda_{2 3}\}$ and the second one $\{ \omega_{2} = \tilde{\omega_2} \}$ with:
\begin{equation}
 \tilde{\omega}_2^{2} = \omega^2 + \frac{2 \lambda_{1 3}(\lambda_{1 2} + \lambda_{2 3} -\lambda_{1 3}) - 
 2 \lambda_{1 2}\lambda_{2 3}}{
 \lambda_{1 2} + \lambda_{2 3} - 2\lambda_{1 3}}
\end{equation}
Both solutions can be simultaneously fulfilled too. In this case we have that: 
\begin{equation}
 \lambda_{1 2} = \lambda_{2 3} \equiv \lambda ~~;~~ \lambda = \omega^2 - \omega_2^2 - \lambda_{1 3} \equiv 
 \tilde{\lambda_{0}}
\end{equation} 
which satisfies Eqs.(\ref{2modesA},\ref{2modesB}) and is thus a condition for a two-modes NS. These three
situations correspond respectively to configurations in  Fig. \ref{figure1}a ($\lambda_{1 2}=\lambda_{2 3}$), 
Fig. \ref{figure1}b ($\omega_{2}^2 = \tilde{\omega_2}^2$) and  Fig. \ref{figure1}f ($\lambda =\tilde{\lambda_{0}}$). It is
worth noting that configuration in Fig. \ref{figure1}a is valid also for the closed  $(\lambda_{1 3} \neq 0)$ case,
as well as Fig. \ref{figure1}b for the open chain (when $\lambda_{1 3} = 0$).

On the other hand, by assuming two equal couplings we have three different solutions: $\lambda = \{ 0, \tilde{\lambda}_{\pm} \}$ , where 
the first one is trivial and accounts for a separated pair of coupled oscillators and a non-coupled third. The second solution allows for 
situations Fig. \ref{figure1}c and  Fig. \ref{figure1}d, defined by:
\begin{equation}
\label{lambda0}
 \tilde{\lambda}_{\pm} =  \lambda_{1 3} \pm \sqrt{(\omega_2^2 - \omega_1^2) (\omega_2^2 - \omega_3^2)}
\end{equation}
Finally for $\{ \omega_1 = \omega_3  ~\wedge~ \lambda_{1 3} = 0 ~\wedge~ \lambda = \omega^2 - \omega_2^2 = \tilde{\lambda_0}  \}$ we have a 
two-mode NS solution (Fig. \ref{figure1}e). 

The presence of one or two non dissipating normal modes prevents full thermalization of the system and
leads to an asymptotic state whose features are analyzed in the following, focusing on entanglement and on quantum
synchronization between the oscillators.

\subsection{Asymptotic Entanglement} 
\label{asymptotic_ent}

When a NS is enabled, decoherence can be prevented in the system leading to asymptotic entanglement that would be
absent in the thermal state. As a measure of entanglement, we will use the well known logarithmic negativity 
which is computable for bipartite Gaussian states \cite{gaussian_entanglement1,gaussian_entanglement2} 
as is our case:
\begin{equation}\label{En}
 E_{\cal{N}} = \max \{0,- \log \nu_{-}\}
\end{equation}
where $\nu_{-}$ is the minimum symplectic eigenvalue of the partial transposed covariance matrix $\tilde{V}_{A B}$, corresponding to 
time reflection of one party. With the help of general expressions we can calculate analytically the asymptotic entanglement when a NS 
is produced. 
Here we present our results for the open chain with equal couplings $(\lambda_{1 2} = \lambda_{2 3} \equiv
\lambda)$ and same frequencies  for the external oscillators $(\omega_1 = \omega_3 \equiv \omega)$, when only one
of the three normal modes is not subjected to dissipation (Fig. \ref{figure1}a) and when only one of them is
dissipating (Fig. \ref{figure1}e) by imposing $\lambda = \tilde{\lambda}_0$. The details of the calculations 
are reported in Appendix \ref{appA}. As initial condition for the natural oscillators we choose a squeezed separable 
vacuum state given by:
\begin{eqnarray} 
\label{initial_condition_original}
\langle q_{i}^2 (0) \rangle = \frac{e^{-2 r_{i}}}{2 ~\omega_{i}} ~~;~~ \langle p_{i}^2 (0) \rangle = \frac{\omega_{i}~ e^{2 r_{i}}}{2}
\end{eqnarray}
where any other first order or second order moments are zero. 

\subsubsection{One mode NS}

As a paradigmatic example of the case in which there is one frozen  normal mode, let us consider the configuration 
given in Fig. \ref{figure1}a. As for the initial condition,  $\omega_1 = \omega_3 \equiv \omega$ in 
Eq. (\ref{initial_condition_original}) and we will assume the same squeezing factor for the external pair, 
i.e. $r_1 = r_3 \equiv r$ while the squeezing in the central oscillator $r_2$ will be irrelevant.

Normal modes coupled to the environment will reach an asymptotic thermal state whose variances are given in
Eq. (\ref{asymptotic}) while  the uncoupled one evolves freely. The asymptotic covariance matrix of the external
oscillators is obtained by expressing the second order  moments of natural oscillators in terms of the normal
modes. Then we substitute respectively the asymptotic expressions corresponding to  the frozen mode (not coupled to
the bath) or the thermalized ones. This yields the following analytical expression for the entanglement:
\begin{equation} \label{entan_1mode}
 E_{\cal{N}} = \max\{0, E_0 + \Delta E (1 + \cos(2 \omega t))\}
\end{equation}
that is defined by a minimum value $E_0$ and an oscillatory term with amplitude $\Delta E$ and frequency $2\omega$:
\begin{eqnarray}
\label{ent_parameters}
 E_0 = &\left\lbrace {\begin{array}{cc}
                            r - r_0^{+}  ~~~  &$for$ ~~~ r \geq 2 r_c  \\
			     r_0^{-} - r  ~~~  &$for$ ~~~ r < 2 r_c
                           \end{array}}\right\rbrace \\
 \Delta E = &\left\lbrace {\begin{array}{cc}
                           2 r_c  ~~~ &$for$ ~~~ r \geq 2 r_c  \\
			    2 r  ~~~&$for$ ~~~ r < 2 r_c
                           \end{array}}\right\rbrace
\end{eqnarray}
where $r_c = (r_0^{+} + r_0^{-})/4$ and the critical values are defined by the following expressions:
\begin{eqnarray}
 r_0^{+} &= \frac{1}{2}\log(4 \lambda^2 \sigma_Q ) \\
 r_0^{-} &=- \frac{1}{2}\log(4 \lambda^2 \sigma_P ) 
\end{eqnarray}
Coefficients $\sigma_P$ and $\sigma_Q$ depend both on the bath's temperature and on the system parameters via the shapes and frequencies of 
the dissipative normal modes as can be seen in their definition in Eq. (\ref{sigmas}). Note that while decoupling of normal modes from the 
bath is temperature independent, the amount of entanglement generated depends on it via the thermalized degrees of freedom.

\begin{figure}[b]
 \includegraphics[width=4.25cm]{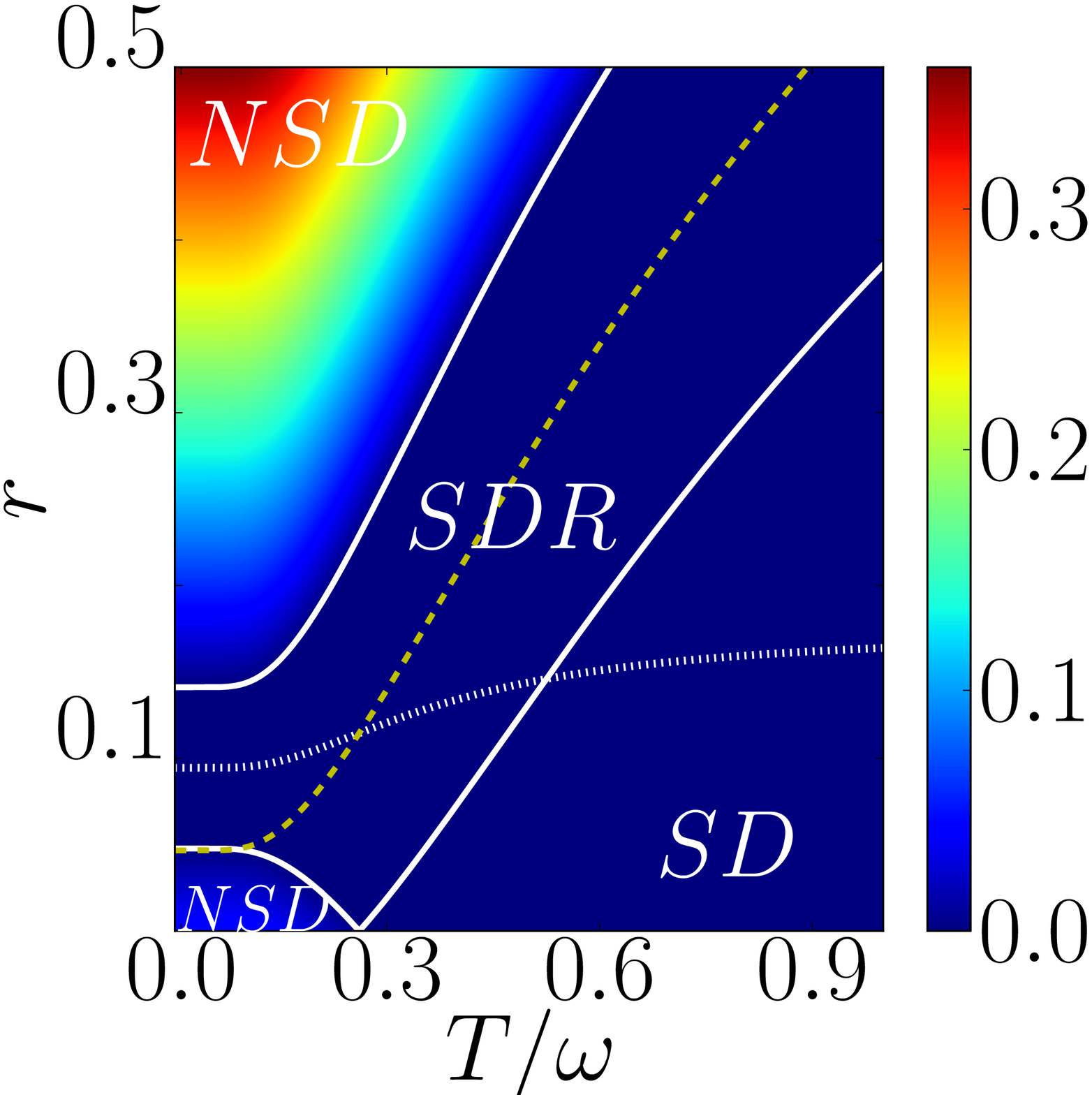}
  \includegraphics[width=4.25cm]{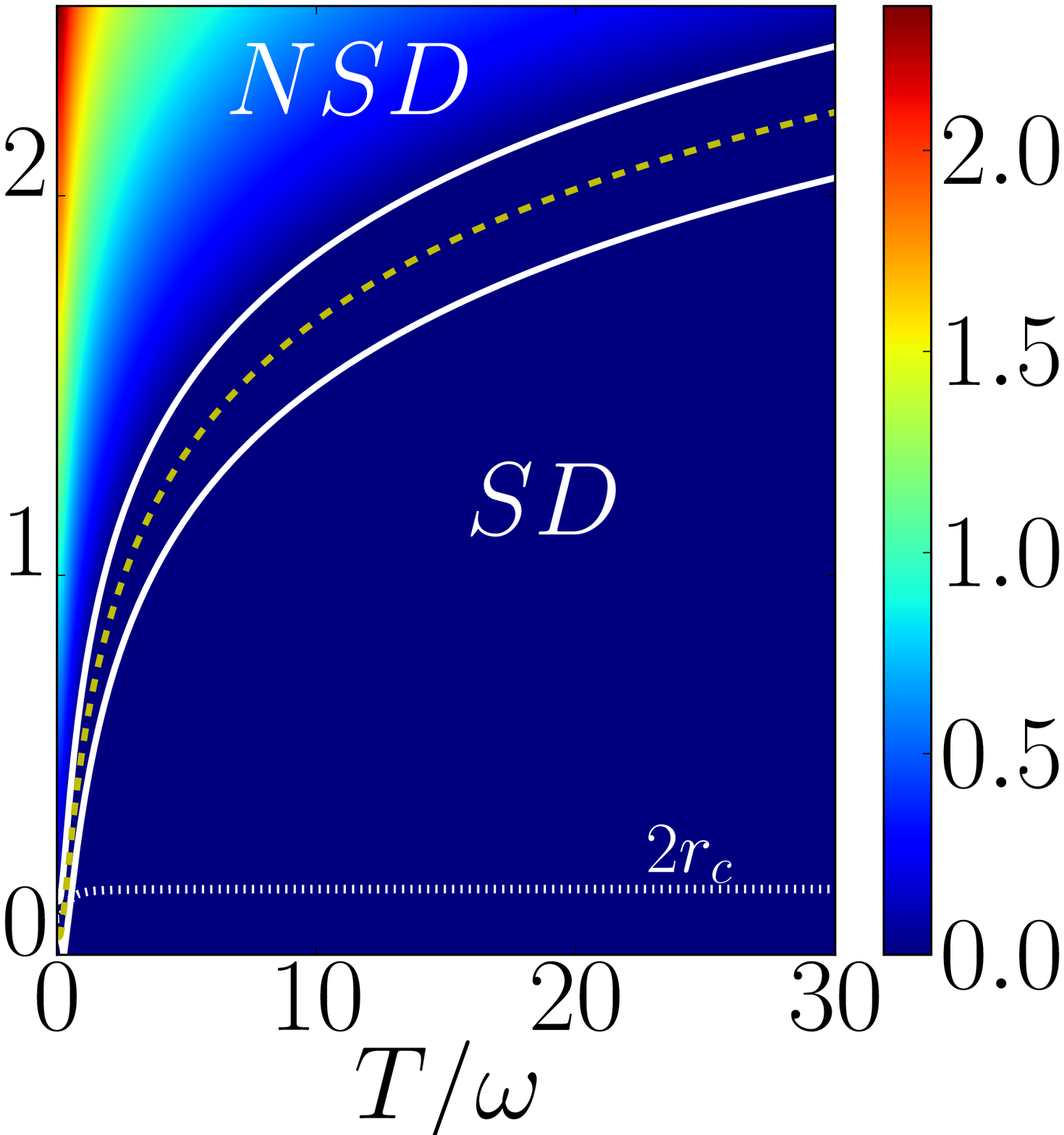}
 \caption{(Color online) Minimum entanglement $E_0$ generated in the asymptotic limit between external oscillators of the chain in configuration in Fig. \ref{figure1}a for low temperatures (left panel) 
 and high temperatures (right panel). The different phases (SD, SDR and NSD) are bounded by the two critical values $r_0^{\pm}$ (separating NSD phase from 
 SDR phase) and $-r_0^{-}$ (separating SDR from SD) that are represented by continuous white lines. Dotted line corresponds to $2 r_c$ and the dashed colored 
 one with $(r_0^{+} - r_0^{-})/2$. We have set $\omega_2 = 1.2 \omega$ and $\lambda = 0.6 \omega$.\label{figure2}}
\end{figure}

The presence of asymptotic entanglement between the external pair of oscillators
in a symmetric chain (independent of the frequency of the central one, but depending on the temperature and initial squeezing)
are illustrated in Figure \ref{figure2}. The minimum entanglement $E_0$ is plotted both for low (left panel) 
and high temperatures (right panel) in the relevant squeezing ranges. Different regions are distinguished in the map and are 
labeled following Paz and Roncaglia notation in Ref. \cite{Paz}: 
sudden death is reached (SD), the asymptotic state consisting of an infinite sequence of sudden 
death and revivals (SDR) and finally when non-zero entanglement is present at all times (no sudden death, NSD). 

An asymptotic entangled state with strictly $E_{\cal{N}} > 0$, can be generated both when $r >
r_0^{+}~( > 2 r_c)$ or equivalently when  $r < r_0^{-}~( < 2 r_c)$ with different origins. In the
first case $(r > r_0^{+})$ the entanglement oscillates between $r-r_0^+$ and $r+r_0^-$ and the
initial squeezing in the natural  oscillators is employed as a resource to generate an entangled
state, while the bath contribution $r_0^{+}$ is a source for its degradation. It's
interesting to see that $r_0^{+}$ is strongly dependent on the bath's temperature while the system
parameters play a  secondary role, only important at low temperatures. Indeed when temperature
increases $(T \gg \omega)$ sudden death of entanglement  can be only avoided by increasing $r$ to
be greater than $r_0^{+} \rightarrow \frac{1}{2} \log \left(4 \lambda^2 T \omega  \left(
\frac{c_{+}^2}{\Omega_{+}^2} + \frac{c_{-}^2}{\Omega_{-}^2} \right) \right) = \frac{1}{2}\log(T) +
ct$. On the other hand the amplitude  of the oscillations in this case is $\Delta E = 2 r_c$ that
has a very weak dependence on temperature quickly reaching a constant value  $\Delta E \rightarrow
\frac{1}{4} \log \left( \omega^2 \left( \frac{c_{+}^2}{\Omega_{+}^2} + \frac{c_{-}^2}{\Omega_{-}^2}
\right)  (c_{+}^2 + c_{-}^2)^{-1} \right)$ when increasing the temperature.

The second case $(r < r_0^{-})$ is only present at low temperatures (of order $0.1\omega$). Here
entanglement oscillates around  $r_0^-+r$ with amplitude $2r$. This means that introducing no
squeezing in the initial state leads to a constant entanglement at $r_0^-$, while adding squeezing
(a resource in the former case) makes entanglement  tend to a SDR phase by widening its oscillatory
amplitude. The fact that thermalization can lead to entanglement at low temperatures is well known
\cite{anders}.

Finally, we can relate critical values $r_0^{\pm}$ with the uncertainty induced by the environment in the virtual oscillator $\tilde{q} = 
(q_1 + q_3)/\sqrt{2}$ position and momenta:
\begin{eqnarray}
 \langle \tilde{q}^2 \rangle_{th} &= \frac{e^{2 r_0^{+}}}{2 \omega} \\
 \langle \tilde{p}^2 \rangle_{th} &= \frac{\omega e^{-2 r_0^{-}}}{2}
\end{eqnarray}
this reveals that when $r_0^{-}>0$ a squeezing in momentum is generated $(\Delta \tilde{p} < \frac{\omega}{2})$ yielding entanglement 
as we have commented above. However note that we have never a minimum uncertainty state with $r_0^{+} >r_0^{-}$ for all temperatures and physical regimes
of system parameters. Indeed the uncertainty relation can be expressed for the virtual oscillator $\tilde{q}$ as $\Delta \tilde{q} \Delta \tilde{p}= 
\frac{1}{2} e^{r_0^{+} - r_0^{-}} > \frac{1}{2}$. The quantity $r_c =$ $(r_0^{+} + r_0^{-})/4$ can be also related with virtual oscillator uncertainties in 
position and momenta as $e^{-r_c} = \frac{\Delta \tilde{p}}{\omega \Delta \tilde{q}} < 1$.

In the left panel of Figure \ref{figure2} we can see the two regions in which $E_0 > 0$ (NSD phases): the big one at the left top corner corresponding to entanglement 
generation by using the initial squeezing in the external oscillators as a resource (once $r>r_0^{+}$) and the small left bottom island that represents 
the environment yielding entanglement via the squeezing generated in the $\tilde{q}$ coordinate when $r < r_0^{-}$. The SDR phase is centered around $2 r_c$ 
(white dotted line) for low temperatures and their amplitude is given by the separation of the dashed colored line $(r_0^{+} - r_0^{-})/2$ from the zero 
squeezing axis. For temperatures greater than that for which $r_0^{-}=0$ (cross point between the dotted and dashed lines) they interchange their roles 
acting now $(r_0^{+} - r_0^{-})/2$ (dashed colored line) as the center of the SDR region and $2 r_c$ as the amplitude. The SD phase is bounded by the quantity 
$r_0^{+} - 4 r_c = -r_0^{-}$ corresponding to the case in which $E_0 + 2\Delta E < 0$ thus no entanglement is present in the asymptotic limit. For high 
temperatures (right panel) we can see how $2 r_c$ reaches a constant value while the pronounced curvature in the SDR region reveals that we can always 
obtain robust entanglement by increasing the squeezing parameter $r$ logarithmically with temperature.

We have to point out that our results resemble those obtained in Ref. \cite{Paz} for two resonant
harmonic oscillators. There a similar entanglement phases diagram has been found and the same
two different mechanism for entanglement generation appear. In that context, both oscillations and the
appearance of the low temperatures NSD phase were attributed to non-markovian effects, while here follow
by considering a final asymptotic (Gibbsian) state  for the normal modes coupled to the bath (that can be
reproduced by a Markovian Lindbland dynamics as is pointed in appendix \ref{appB}). Presence of 
a third oscillator in the system allows for manipulation of the width of entanglement phases at low temperatures 
(specially the low squeezings NSD one) by tuning the free system parameters $\omega_2$ and $\lambda$.

\subsubsection{Two modes NS}

Let us now consider the case in which two modes become decoupled from the bath and in particular focusing
on the configuration in Fig. \ref{figure1}e. This is indeed a symmetric open chain configuration as before, but now we have a
special value of the couplings $\lambda =\tilde{\lambda}_0$ leading to a larger NS. The calculation is similar to the previous  
one while now we have that only one of the normal modes thermalizes and the other two have a free evolution decoupled from the bath. 
This leads to a less compact expression for asymptotic entanglement and more details are reported in Appendix \ref{appA}. Still,
a similar phase diagram for entanglement can be found in this case by numerical evaluation of logarithmic negativity from Eq. (\ref{nu_two}) 
given in the appendix \ref{appA}. Results are shown in Fig. \ref{figure3} in the same range of 
squeezing and temperatures as in the previous (one mode NS) case. For low temperatures (left panel) the low temperatures low squeezing NSD 
island of Fig. \ref{figure2} that corresponds to the environment acting as a resource for entanglement generation disappears, since the bigger 
one expands to low squeezing. Degradation of resources by the environmental action here is not sufficient to prevent entanglement production 
even in the non-squeezed ($r=0$) case for $T < T_c$ since actually the mode $\epsilon$ is also contributing to entanglement generation. On the other 
hand the entanglement phases shows the same structure for high temperatures (right panel) where the only difference resides in the attenuated growth 
for entanglement when $r$ increases (see color bars).

\begin{figure}[b]
 \includegraphics[width=4.25cm]{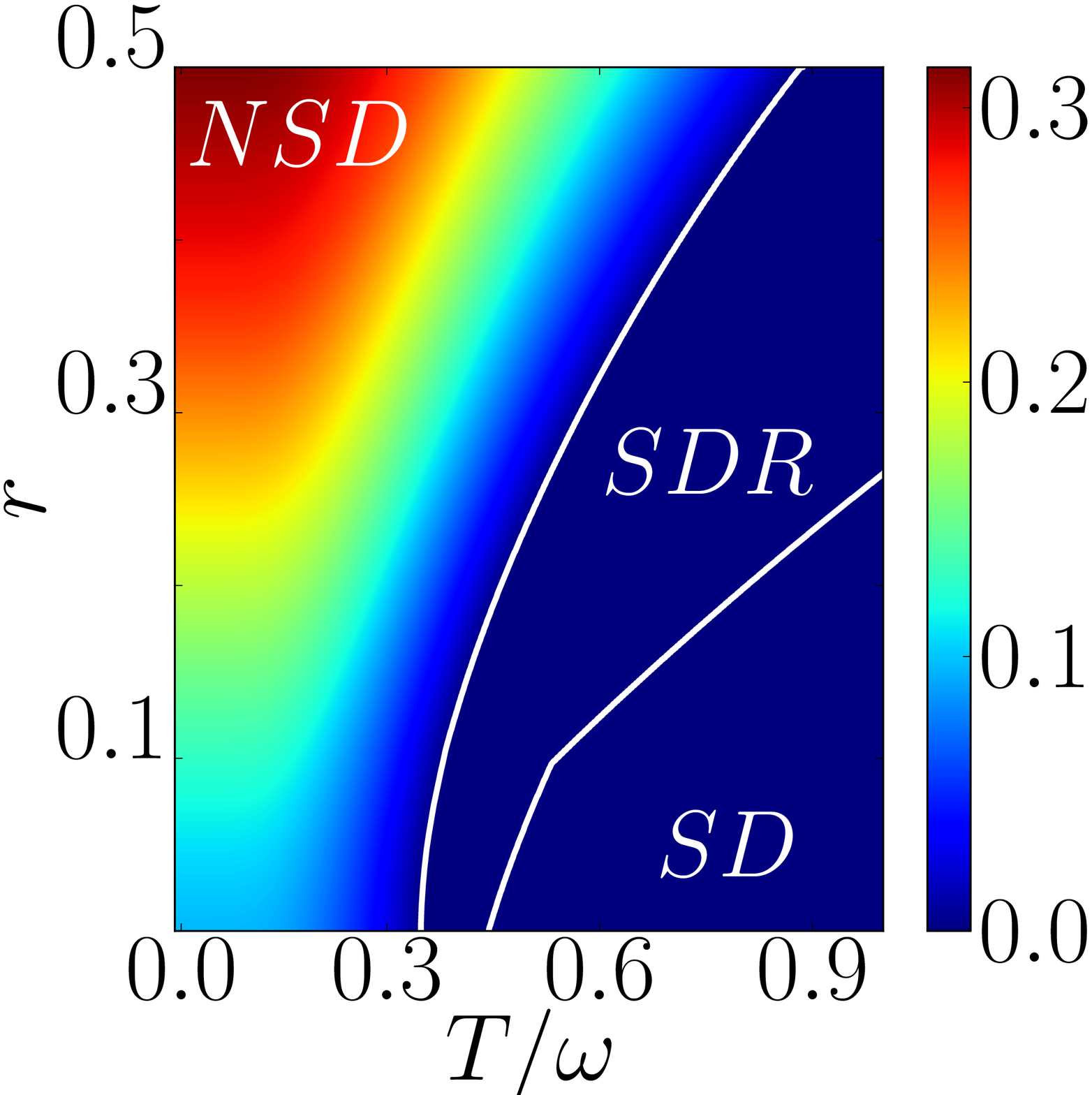}
  \includegraphics[width=4.25cm]{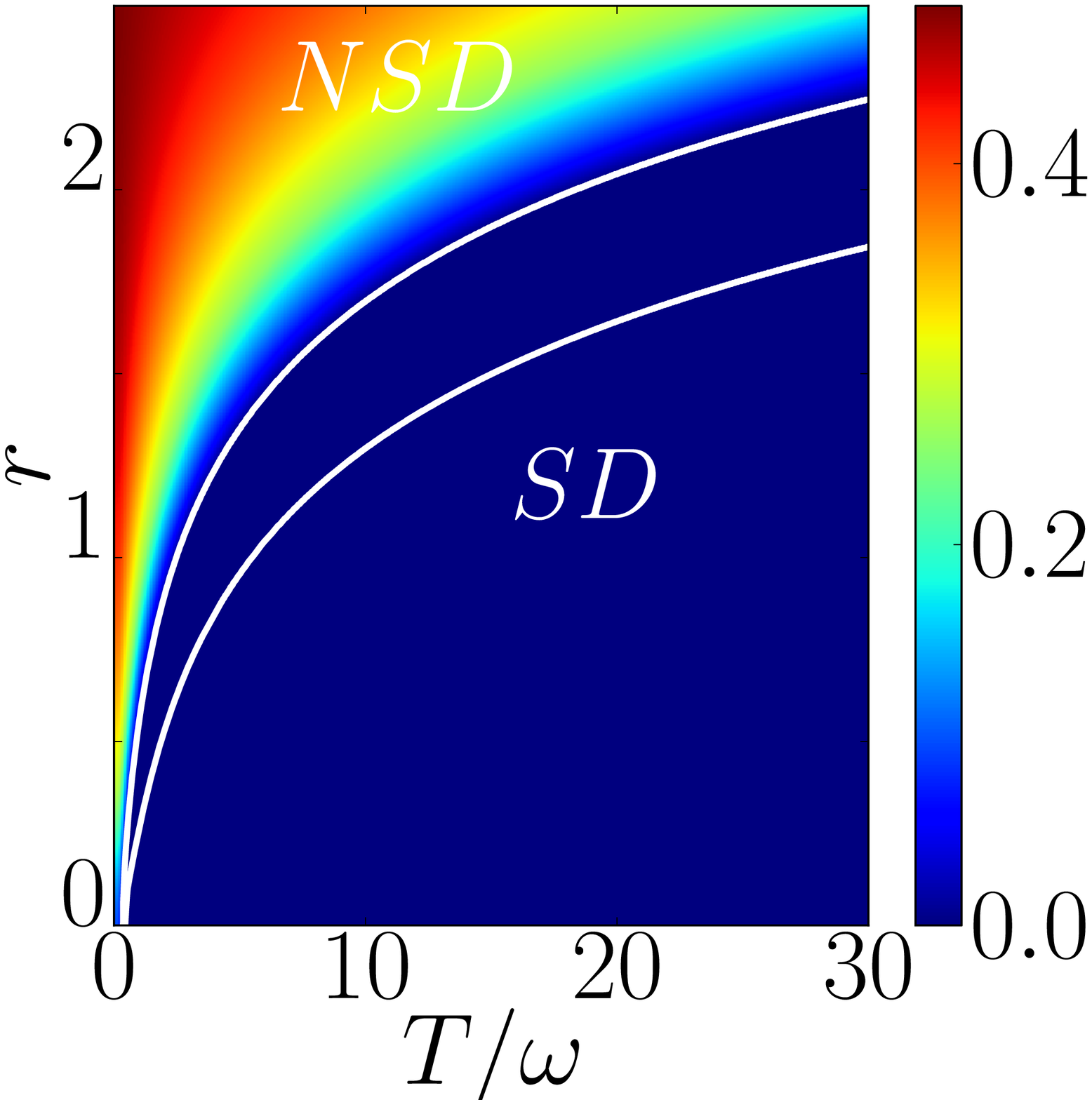}
 \caption{(Color online) Minimum entanglement in the asymptotic limit between external oscillators of the chain in the two modes NS configuration in 
 Fig. \ref{figure1}e for low temperatures (left panel) and high temperatures (right panel). The different phases (SD, SDR and NSD) are 
 bounded by the continuous white lines obtained by numerical evaluation. We have set $\omega_2 = 1.2 \omega$.\label{figure3}}
\end{figure}

{\it{Notice}} that all the expressions have been calculated in the limit of small gamma assuming a final Gibbs state for decohered eigenmodes and free 
evolution for nondecohered ones. Of course this situation can only be perturbative, since for stronger coupling to the bath the eigenmodes become 
increasingly coupled among them, through second order processes mediated by the bath. This necessarily leads to decoherence of all eigenmodes, at 
a low rate though.

\subsection{Quantum synchronization}  \label{sync} 

In this section we analyze the dynamics of the system showing the existence of a parameter manifold where
the oscillators oscillate in phase, synchronously, in spite of having different natural frequencies. Full
dynamics for Gaussian states is characterized by  mean values of position, momenta and variances (second order
moments). The possibility to have synchronization in this system is important for two reasons:  (i) this
phenomenon has been largely studied in non-linear systems but we show that for dissipation in a common bath, 
it can arise even among harmonic oscillators;  (ii) few attempts have been done to extend it to the quantum
regime and we show here that this phenomenon witnesses the presence of robust quantum correlations, being
actually a synchronous steady state accompanied by asymptotic entanglement.

Spontaneous  synchronization -where no external forcing is considered- has been recently studied in some quantum 
systems like optomechanical cells or 
nanomechanical resonators \cite{sync1,our,sync2,syncnet}. Actually the phenomenon of mutual synchronization has
been  extended to the quantum regime considering a different scenario in Ref. \cite{our} where two coupled
harmonic  oscillators with different frequencies and initially out of equilibrium are studied during their
relaxation towards a thermal state.  Synchronization was  reported in first and second order moments during a
long transient and accompanied by the robust preservation of quantum correlations (quantum discord
\cite{discord1,discord2}) between  oscillators. Two oscillators dissipating in a common bath are actually
preserving asymptotically their entanglement and retaining a larger energy than in the thermal state only if
they are identical \cite{Galve}. In this (symmetric) case they also evolve towards a synchronous asymptotic
state \cite{our}. 
When three elements are considered, we have shown above that the symmetric chain can reach
an asymptotic regime with entanglement between the external oscillators, independently on the frequency of the
central one.Then asymptotic synchronization between the external pair is also expected. Beyond this symmetric case,  
more interesting is the possibility offered by a chain to freeze all the oscillators out
of the thermal state when  their frequencies are all different, as discussed below.

The long time dynamics of our system can be straightforwardly calculated by assuming that normal modes  which
are coupled to the bath get thermalized, while uncoupled ones have a free evolution. This is sufficient to
analyze the presence of synchronization  in the asymptotic state. In general quantum mutual synchronization
appears always in one-mode NSs among natural oscillators linked by the non-dissipative mode, as long  as they
have an asymptotic dynamics with only one oscillatory contribution. Phase or antiphase synchronization at the
non-dissipating normal frequency is possible in first order moments depending on the sign of their
$\mathcal{F}$ matrix coefficients, while only in-phase synchronization  occurs for second order moments at
twice the frequency of first order ones. Let us illustrate it in some situations and  compare with the time
evolution of $\langle q_i^2 \rangle ~\forall i~$ (Fig. \ref{figure_moments}) when considering a simple 
Markovian dynamics in the weak coupling limit (see Appendix \ref{appB}).

\begin{figure}[b]
\includegraphics[width=9cm]{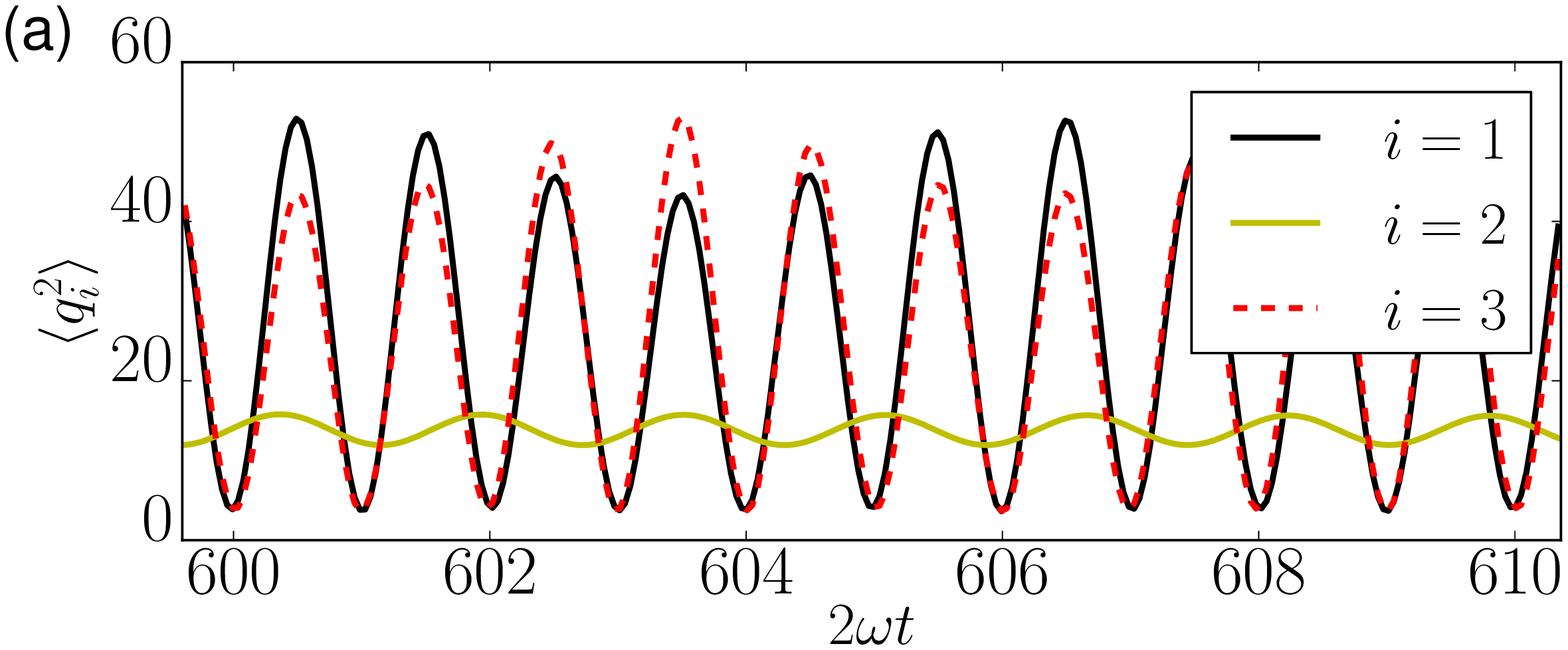}
\includegraphics[width=9cm]{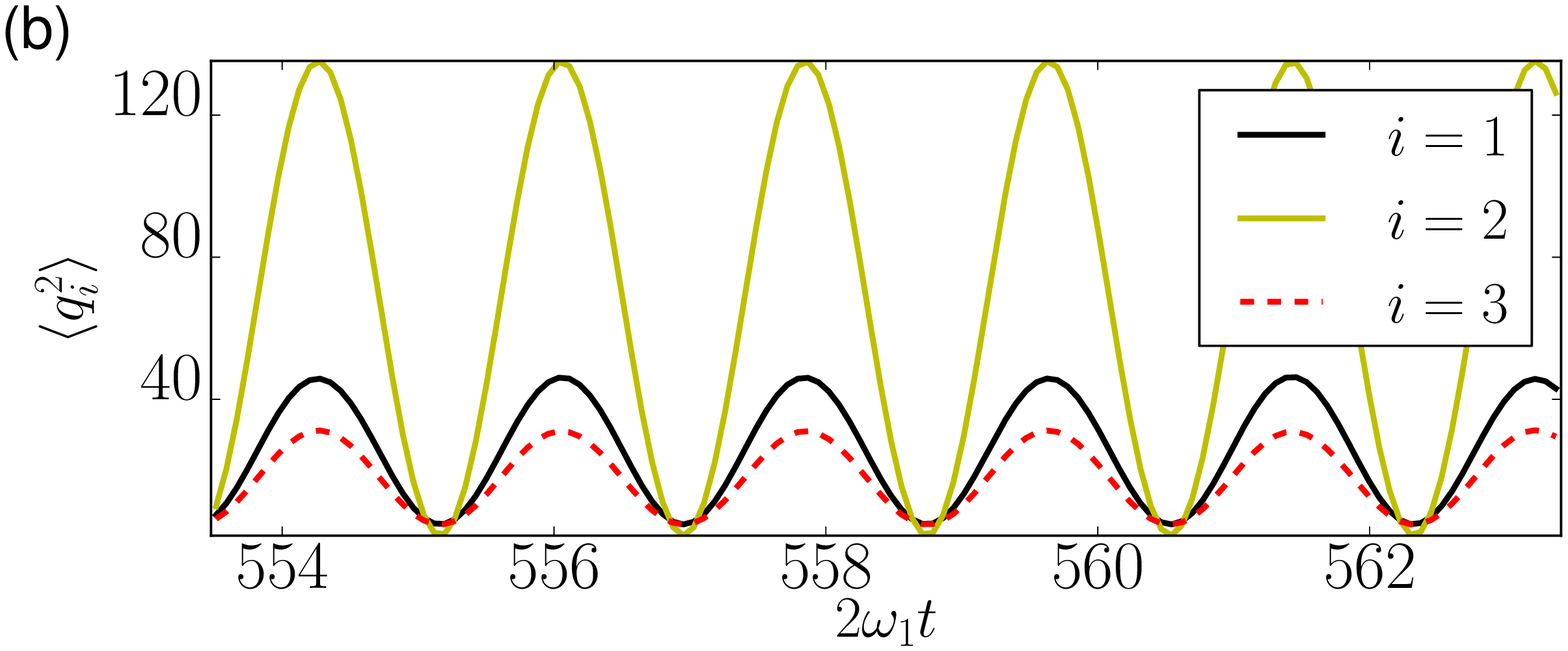}
\includegraphics[width=9cm]{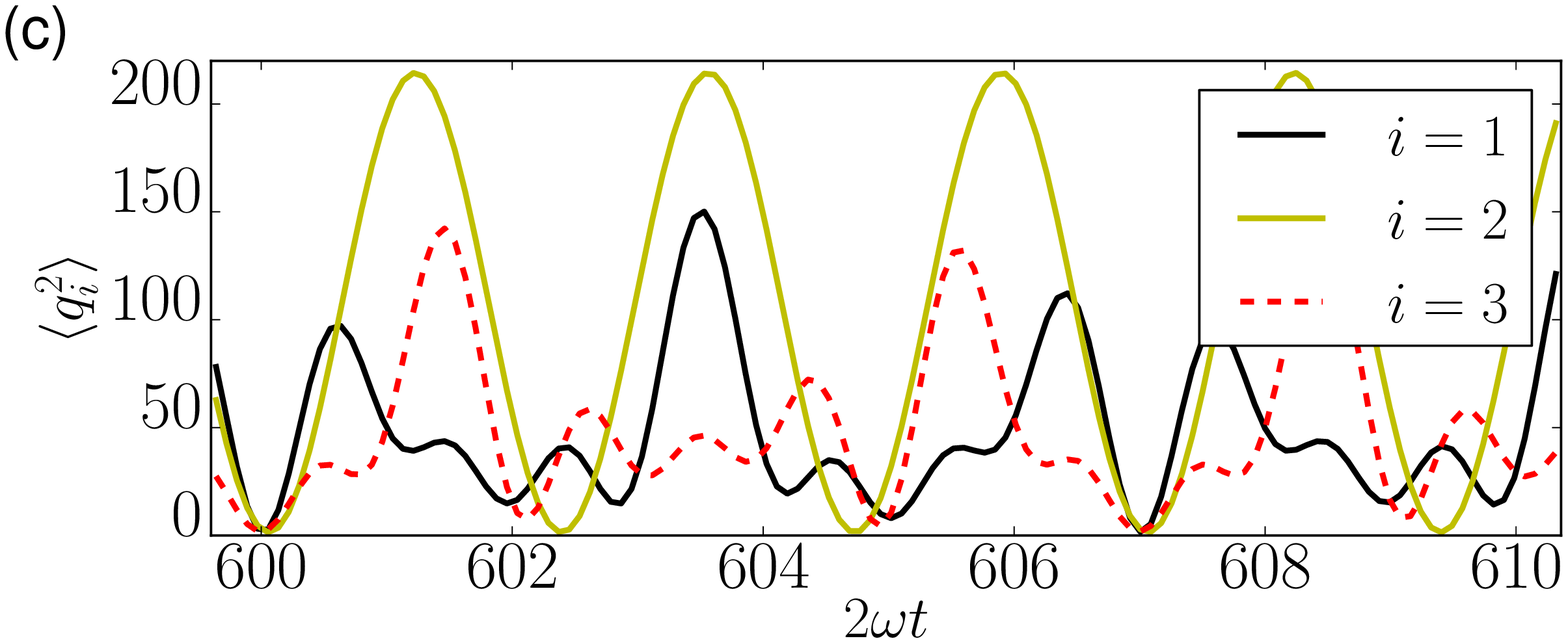}
\caption{(Color online) Evolution of position variances for each oscillator in the open chain (see legend) for 
(a) configuration in Fig. \ref{figure1}a where one mode NS is generated $(\omega= 1.3\omega_2 ~,~ \lambda=0.4 \omega_2^2)$ synchronizing the external oscillators at 
$2 \omega$, (b) configuration in Fig. \ref{figure1}c where a different mode NS is generated $(\omega_1 = 1.2\omega_2 ~,~ \omega_3=1.3 \omega_2 ~,~ \lambda=0.4 \omega_2^2)$ 
producing synchronization in all pairs of oscillators at $2\Omega_\epsilon$ and (c) configuration in Fig. \ref{figure1}e when a two modes NS 
is generated ($\omega= 1.3\omega_2 ~,~ \lambda=\tilde{\lambda}_0$) and synchronization is lost. Bath parameters for the simulation are in 
all cases $T=10\omega_2$, $\gamma = 0.07 \omega_2$ and $\Lambda= 50 \omega_2$. 
\label{figure_moments}}
\end{figure}

Consider firstly the specific case of an open chain with equal couplings 
and frequencies in the external oscillators (corresponding to situation in Fig. \ref{figure1}a). The form of the non-dissipative normal mode is 
${\bf C}_\delta = (1 , 0, -1)^T/\sqrt{2}$ so synchronization will emerge only between external oscillators in antiphase for the position and  
momenta at frequency $\Omega_\delta = \omega$ (the normal mode frequency) and for the second order moments (necessarily in-phase 
and at $2\omega$). The central oscillator instead decays into the thermal state, its initial oscillations being suppressed in the 
long time dynamics. This case is shown in Fig. \ref{figure_moments}a, where synchronization appears after a transient only for the external 
oscillators of the open chain while the central oscillator looses oscillation amplitude. 

In the latter case synchronization appears between identical unlinked ($\lambda_{13}$) oscillators in a symmetric chain 
(Fig. \ref{figure1}a). More peculiar is the case in which all oscillators have different frequencies and eventually couplings. 
In the case of Fig. \ref{figure1}c we actually have that the non-dissipative mode involves all the three
oscillators ${\bf C}_\epsilon = c_\epsilon (\omega_3^2-\omega_2^2~,~ \omega_3^2-\omega_2^2 - \tilde{\lambda}_{\pm}~,~ 
\tilde{\lambda}_{\pm})^T$ with $\Omega_\epsilon = \sqrt{\omega_2^2 + \tilde{\lambda}_{\pm}}$. This can actually give rise 
to  synchronous dynamics of all the oscillators, in spite of the difference in their natural frequencies. Since 
one of the components has different sign than the other two in ${\bf C}_\epsilon$ two of the oscillators first moments 
will synchronize in-phase between them and in antiphase with the third one. In Fig. \ref{figure_moments}b a total 
synchronization is produced involving all three (different) oscillators, consistently with the fact that the non 
dissipative normal mode  $\epsilon$ involves all three oscillators.

A different situation is produced when we have a two-modes NSs since two oscillatory contributions are present
in the asymptotic limit of the  natural oscillators. Here synchronization is only possible when the two normal
modes frequencies are the same. An example is the open chain
with a two-modes NS (Fig. \ref{figure1}e) where apart from the previous non-dissipative  mode ${\bf C}_\delta$,
actually the mode ${\bf C}_\epsilon = (1, -2, 1)^T /\sqrt{6}$ with frequency $\Omega_\epsilon =
\sqrt{2\omega_2^2 -  \omega^2}$ does not dissipate either. In this case synchronization is destroyed by the
presence of mode $\epsilon$ and it can be only  recovered when $\Omega_\epsilon$ equals $\Omega_\delta$, i.e. in
the trivial case of independent $(\lambda=0)$ identical oscillators  $(\omega_2 = \omega)$. Lack of synchronization 
as well as a multimode oscillation are shown in Fig. \ref{figure_moments}c.

The initial state employed for simulations is a squeezed separable vacuum state, where the squeezing parameters have
been chosen to be different $(r_1= 2, r_2= 2.5$ and $r_3= 3)$. In general,  we have tried to avoid special
initial conditions that could have filtered just one normal mode into the dynamics. What we discussed is
therefore  the emergence of synchronization as a dynamical process when considering more general  initial
states, leading  to robust conclusions.

The scenarios here discussed allow to establish the effect of having one or two modes NS in the
configurations of open chains (Fig. \ref{figure1}a and e). The same analysis can be extended to other
cases where a different normal  mode is uncoupled from the environment. For instance, the
configurations in  Fig. \ref{figure1}c and d admit only one non-dissipative normal mode that involves the three
oscillators, producing then a collective synchronization of the chain.

\section{Thermalization and robustness of quantum correlations} 
\label{robustness}

Creation of NSs is a powerful tool to avoid decoherence and produce synchronized dynamics as we have seen in 
previous sections. However, the conditions leading to NSs are satisfied only in some parameter manifolds and it
is relevant to see which is the effect of deviations from these couplings and detunings, that could also arise
from the difficulty of experimental tuning. In this case, dissipation is present in all
normal modes and the effective couplings of Eq.  (\ref{interaction}) are all different from  zero. Thus a thermal 
state is finally reached in all the degrees of freedom. 

In absence of NS entanglement is lost after a finite time. Although the asymptotic state is simply the thermal (Gibbs) state, damping dynamics 
of the normal modes with different decoherence and relaxation time scales is present, producing a rich behaviour in which synchronization or 
high quantum correlations may emerge during a large transient before the final thermalization of the system. These effects 
have been reported recently in the case of two harmonic oscillators, where disparate decay rates between the two normal modes is produced 
for small deviations from the resonat case \cite{our}.

A dynamical description of the system in terms of a Markovian Master Equation (Appendix  \ref{appB})
 reveals the central influence of the effective couplings in the
relaxation time scales of the different normal modes. In  this context, the ratio between the two smallest decay
rates, defined as:
\begin{equation}
R \equiv \frac{\Gamma_{0}}{\Gamma_{1}} = \frac{\kappa_0^2}{\kappa_1^2} 
\end{equation}
provides important information about the dynamics of the system, being in fact one of the central
figures in order to predict the  robustness of correlations between oscillators or the emergence of
synchronization as we will see in the following. In presence of disparate  decay rates ($R<<1$), a large time
interval appears between thermalization of the two modes with largest damping coefficients ({\it strongly-damped} 
modes) and thermalization of the mode with the smallest one ({\it weakly-damped} mode) . This produces the
emergence, after a transient, of a  long interval in which the {\it weakly-damped} mode is effectively the only
one present in the dynamics, hence producing the synchronization between pairs  of oscillators linked by this
normal mode and the slow decay of quantum discord between these pairs. On the other hand, when the decay 
rates are similar ($R \sim 1$) the different modes are present for all times inhibiting synchronization, and the
survival of correlations  associated to one of the modes for long times is lost. These phenomena will be exemplified in
the scenario of an open chain with equal couplings  ($\lambda_{1 2} = \lambda_{2 3} \equiv \lambda $).

\begin{figure}[t]
\includegraphics[width=9cm]{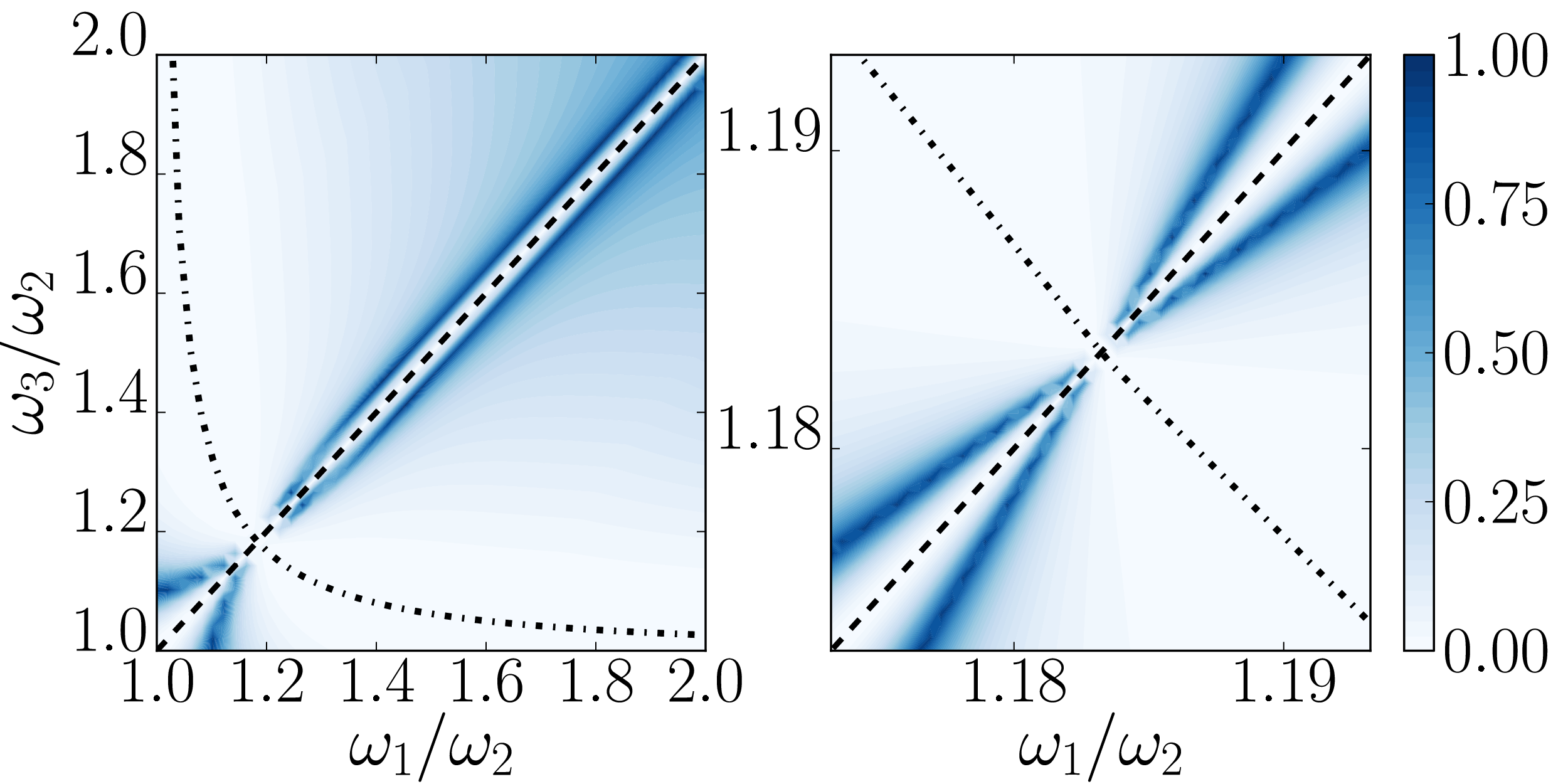}
 \caption{(Color online) Map of $R$ for $\lambda = 0.4 \omega_2^2$ as a function of the open chain frequencies. Dashed and
 dashed-dotted lines represent the  non dissipative parameters manifolds of Figure \ref{figure1}a and c  respectively. The
 right panel is a zoom of the vicinities of the two-modes NS cross point in the left panel .
 \label{figure_kappas}}
\end{figure}

In Figure \ref{figure_kappas} we represent $R$ showing broad regions in which a {\it weakly-damped} mode exists (white
regions) near the NSs manifolds corresponding to the configurations in  Fig. \ref{figure1}a (dashed line),
\ref{figure1}c (dashed-dotted hyperbola)  and \ref{figure1}e (the crossing point). Out of these regions there
is no separation of scales for the decay rates (blue  regions) and all rates become progressively similar. We
point out that the blue region wrapping the diagonal in Fig. \ref{figure_kappas} acts as boundary for the two white ones, since 
a different mode (with  radically different shape) is {\it weakly-damped} in each white region. The coupling $\lambda$ is
related to the position of the dashed-dotted  hyperbola by Eq.  (\ref{lambda0}) and the width of white regions,
making them broader as $\lambda$ increases and tighter when it decreases.

\subsection{Quantum correlations}

Even if out of the NS conditions entanglement suffers a sudden death, following Ref. \cite{our} other quantumness
indicators, such as quantum discord can remain robust in regions where disparate decay rates $(R<<1)$ exist.  
By using an adapted measure of discord for Gaussian bipartite states \cite{gaussian_discord1, gaussian_discord2} we 
observe the existence of a {\it plateau} in the dynamical evolution of discord between
pairs of oscillators which are linked  by a {\it weakly-damped} normal mode. More precisely, in the white region of Figure
\ref{figure_kappas}, close to the dashed-dotted hyperbola, the  {\it weakly-damped} mode links the three natural
oscillators, producing a {\it plateau} in the evolution of discord for all pairs. 
Moving to the tighter  white region, close to the dashed diagonal line, the {\it weakly-damped} mode only involves 
the external oscillators pair of the open chain,  leading to a slowly decaying discord only for this pair of oscillators. In
blue regions no {\it plateau} is observed for discord, reaching  for each pair the value corresponding to the
thermal state in shorter times. 

\begin{figure}[t]
\includegraphics[width=8.5cm]{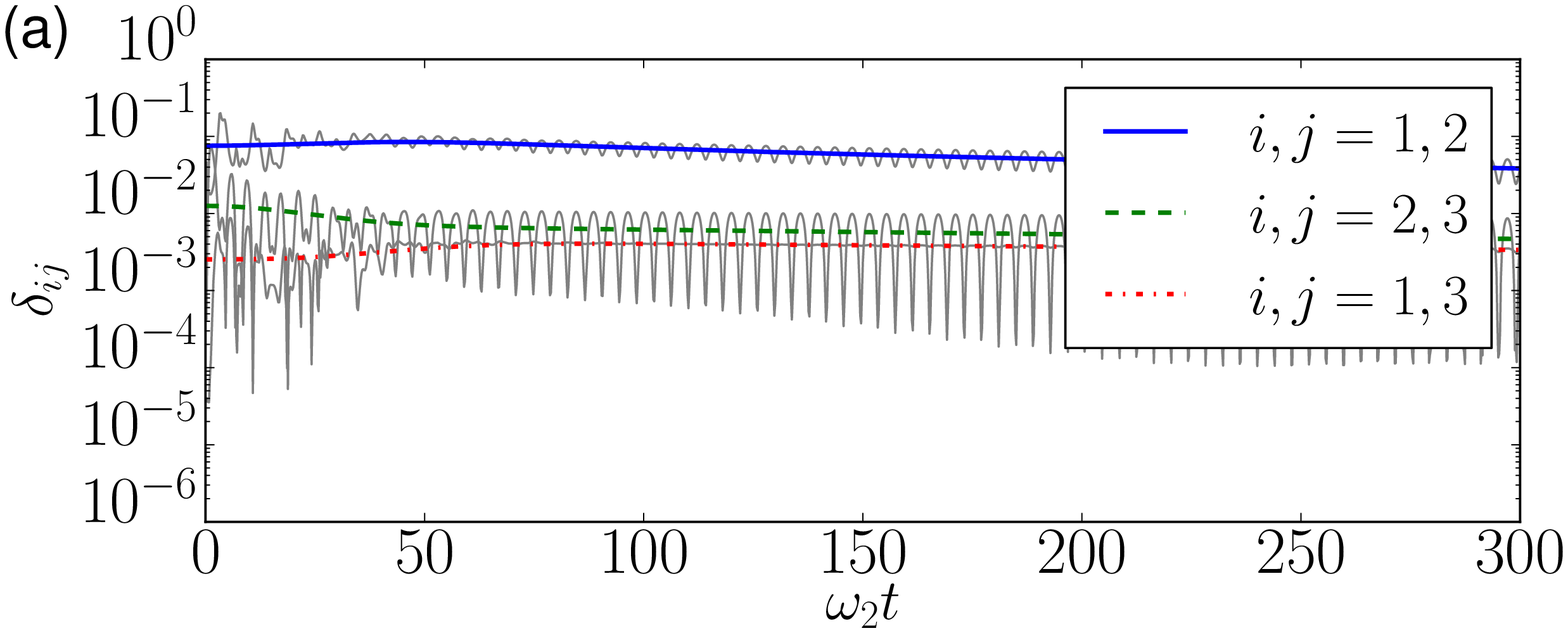}
\includegraphics[width=8.5cm]{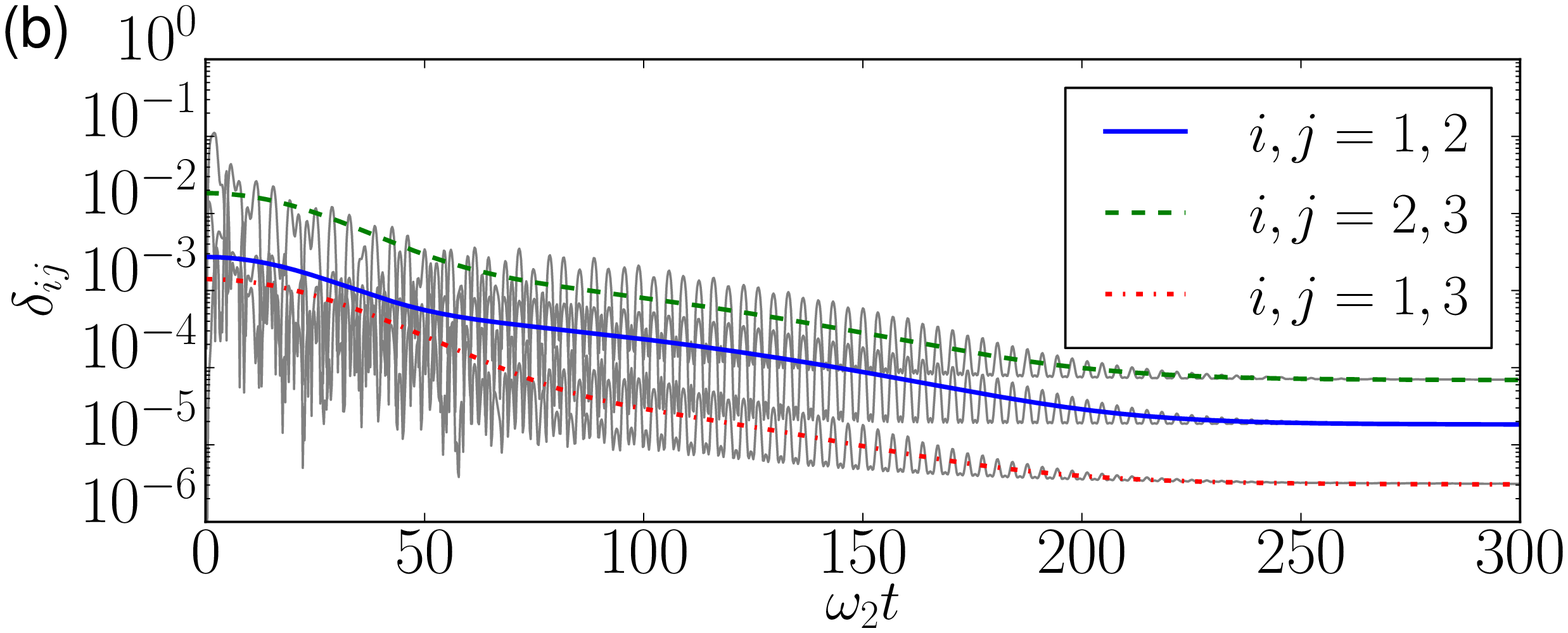}
 \caption{(Color online) Time evolution of discord for the three pairs of oscillators in the open chain (see legend) for two different regions of 
 Figure \ref{figure_kappas}, setting $\lambda = 0.4 \omega_2^2$, $\omega_3 = 1.6\omega_2$ and changing $\omega_1$. (a) Near the 
 dashed-dotted hyperbola $(\omega_1 = 1.1\omega_2)$ and (b) far away from it $(\omega_1 = 1.9\omega_2)$. Bath parameters have been taken 
 for the simulations $T=10\omega_2$, $\gamma=0.07\omega_2$ and cutoff frequency $\Lambda= 50\omega_2$. \label{figure_discord}}
\end{figure}

Figure \ref{figure_discord} shows time evolution of discord in logarithmic scale 
for the three pairs of oscillators (see colors) for a selection of parameters close and far away from the dashed-dotted
hyperbola  (Fig. \ref{figure_discord}a and \ref{figure_discord}b respectively). The initial condition has been taken to 
be a squeezed separable vacuum state with same squeezing parameters as in Fig. \ref{figure_moments} and will be kept for further 
simulations. A Gaussian filter has been employed to eliminate fast oscillations  (original quantities are plotted in gray), 
in order to make it easier to identify the {\it plateau} characterizing discord robustness.

\begin{figure}[t]
\includegraphics[width=8.5cm]{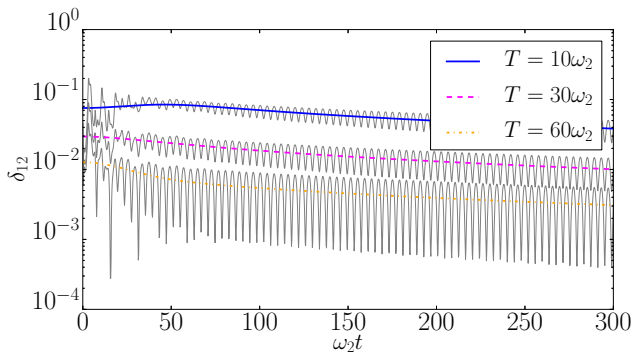}
 \caption{(Color online) Time evolution of discord for a pair of linked oscillators ($1,2$)
 of the open chain for different bath temperatures (see legend).
 We have set $\lambda = 0.4 \omega_2^2$, $\omega_3 = 1.6\omega_2$ and $\omega_1 = 1.1 \omega_2$. The rest of bath parameters has been kept
 $\gamma=0.07\omega_2$ and $\Lambda= 50\omega_2$. \label{figure_discord_T}}
\end{figure}

As already seen for asymptotic entanglement, the effect of increasing the bath's temperature is, in general,
a degradation of quantum effects. It is therefore important to see how robustness of discord is a feature
present also in hotter environment. The main effect when increasing T 
is that the final thermal state   (by virtue of Eq. \ref{asymptotic2}) displays lower correlations so that the amount of 
discord that  can be maintained in a robust way diminishes. In Figure \ref{figure_discord_T} we show the evolution of discord for 
a pair of linked oscillators ($1,2$) for increasing T by factors $3$ and $6$ (other parameters as in Fig. \ref{figure_discord}a): 
while the {\it plateau} is present for all temperatures and their (negative)  slope is very similar, a lower amount of discord 
can be generated in the initial transient producing a shift of the curves to lower  values (oscillations are increased
by the logarithmic scale of the plot). This degradation by temperature effects can be avoided by  increasing the
squeezing in the initial separable vacuum state like in the case of entanglement presented in section
\ref{asymptotic_ent}. 

\subsection{Synchronous thermalization}

With respect to the emergence of synchronization for pairs of oscillators when the NS is lost, we have to point
out that when thermalizing the system reaches a stationary state where oscillations are suppressed. We 
therefore restrict our analysis to a transient (which becomes longer the more we approach one of 
the NS conditions) where oscillations in the first and second order moments are still present in the 
dynamics. In this situation synchronization of first and second order moments can be estimated quantitatively 
by introducing a typical indicator, that is defined for two generic time-dependent functions $h(t)$ and $g(t)$ as:
\begin{equation} \label{C}
C_{h,g}(t,\Delta t)=\frac{\int_t ^{t + \Delta t} dt' (h-\bar{h})(g-\bar{g})}{\sqrt{\Delta h \Delta g}}
\end{equation}
where $\Delta h = \int_t ^{t + \Delta t} dt' (h-\bar{h})^{2} $ and $\bar{h}=\frac{1}{\Delta t} \int_t ^{t + \Delta t} dt' h(t')$. When 
evolutions are phase or antiphase synchronized we will obtain $|C| \sim 1$, while for very different dynamics we will obtain a value of 
$C$ near to zero.

\begin{figure}[t]
\includegraphics[width=9cm]{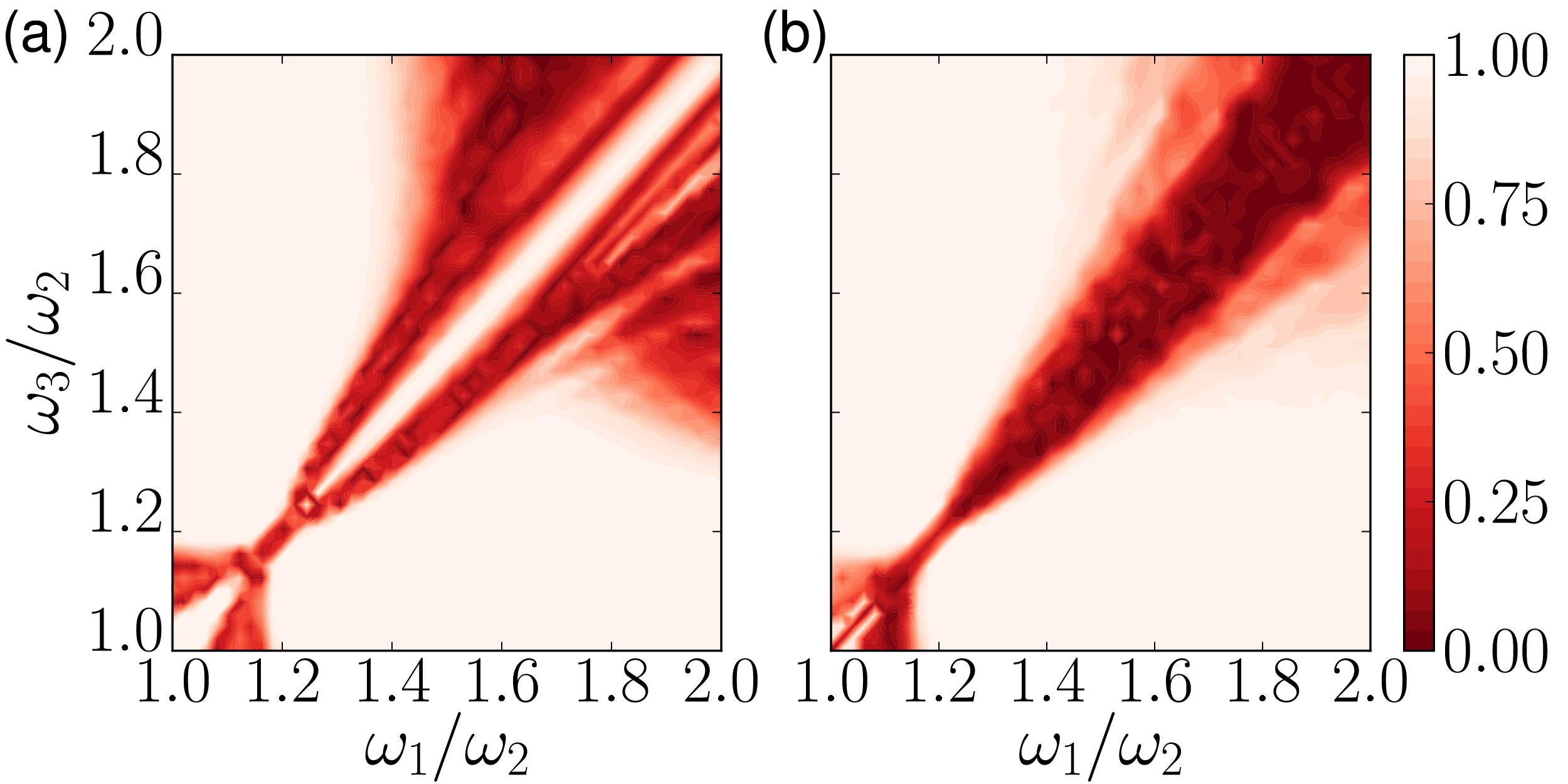}
 \caption{(Color online) Absolute value of the synchronization indicator $|C(t, \Delta t)|$ for  
  position variances $(\langle q_i^2 
 \rangle)$ for (a) the external pair of oscillators $1,3$ and (b) a linked pair $1,2$. 
 The synchronization factor is plotted at time 
 $t=\min \{t_{MAX}, \Gamma_0^{-1}\}$ where $t_{MAX}= 5000\omega_2^{-1}$ 
 (the maximum time used in the simulations) in order to obtain a map 
 in which oscillations were not yet suppressed. 
 We have used $\Delta t = 15 \omega_2^{-1}$ and the same bath parameters as in former figures. \label{sincro_map}}
\end{figure}

Figure  \ref{sincro_map} shows the synchronization indicator  $C_{\langle q_i^2 \rangle, \langle q_j^2
\rangle}$ with position variances of (a) the external pair of oscillators $i,j=1,3$ and (b) for $i,j=1,2$  of
the open chain with identical couplings and varying the external oscillators frequencies (in the same range as in Fig. \ref{figure_kappas}
).  We see immediately the high resemblance with the $R$ map of Fig. \ref{figure_kappas} and some  interesting differences induced by the
shape of the normal modes. Effectively the external pair of oscillators ($1,3$) synchronizes  
 $(C \sim 1)$ in all regions where disparate decay rates $(R \ll 1)$ are predicted since these two oscillators 
are linked by the {\it weakly-damped} normal mode in these regions (Fig. \ref{sincro_map}a). 
As for the internal pair ($1,2$), it does depend on the {\it weakly-damped} mode in the vicinities of the
dashed-dotted hyperbola where synchronization is actually found.
On the other hand,   near the diagonal the {\it weakly-damped} mode approximates to ${\bf C}_{\delta} 
= 1/\sqrt{2}(1,0,-1)$ excluding the central oscillator from the induced collective motion.  Consistently,
the $1,2$ pair (Fig. \ref{sincro_map}b) does not synchronize for $\omega_1\sim\omega_2$ (near the diagonal).

We finally point out that, as expected, the synchronization frequency is that of the  {\it weakly-damped} mode
$(\Omega_0)$ for the first order moments (position and momenta) and $2 \Omega_0$
 for the second order  momenta.

\section{Conclusions} 
\label{conclusions}

 Decoherence in an open quantum system can be avoided or reduced by tuning the
system parameters in a common environment context. The  shape of the interaction
Hamiltonian between system and bath can be used in order to engineer the protection of some
degrees of freedom  from the environmental action. This analysis has been solved in
the present work for a system of three coupled harmonic oscillators in contact
with a bosonic  bath in thermal equilibrium, developing the necessary general
relations so as to obtain a NS composed by one or two non-dissipative normal  modes.

Different open and close chain configurations have been explored, highlighting the richer
variety of NS configurations available when the dissipative system is extended from two to three 
harmonic oscillators.For a symmetric open chain (equal frequencies of the external pair of 
oscillators and same coupling)
a closed analytical expression for the asymptotic entanglement  (logarithmic negativity) 
has been derived observing the appearance of three different phases depending 
on temperature and squeezing of the initial state (sudden death of entanglement, a infinite 
series of sudden death events and revivals and asymptotic robust entanglement). Sudden 
death of entanglement can be avoided for arbitrarily high bath temperatures by increasing 
the squeezing in the initial  state for both one or two modes NS. Remarkably this critical 
squeezing in order to avoid sudden death depends logarithmically on temperature.  

Other dynamical effects such as the emergence of  synchronization of mean
values and variances have been analyzed in different situations by simply
assuming the relaxation to a thermal state of the normal modes coupled to the
environment. Coherent oscillations appear when only a surviving normal mode is 
present in the dynamics, inducing synchronization in the natural oscillators that
depend on it. Interestingly the parameter manifold leading to NSs include several
not symmetric configurations: for instance an hyperbolic relation among frequencies can
be satisfied for identical couplings in an open chain; in this case both asymptotic
entanglement and synchronization are predicted even if all the oscillators natural
frequencies are different (a possibility offered by a chain of three oscillators and
absent in the case of two). A more general scenario where oscillators networks are 
considered with dissipation acting globally and locally has been discussed in Ref.\cite{syncnet}.

Furthermore an analysis of situations in which the NS conditions are not accomplished
at all has been performed. Indeed, important properties 
can be present although when deviating from NS conditions
entanglement does not survive: robust conservation of discord during a long transient 
dynamics and the emergence of synchronous oscillations are found before thermalization.
These effects are interpreted in relation to disparate decay rates for the normal modes. 
As long as there is a {\it weakly-damped} mode surviving among several {\it strongly-damped}
modes, effects such as robust discord and synchronization
arise among the oscillators following this normal mode.

While we have focused along this paper in specific cases of all the possible three
oscillator configurations (in which calculations are greatly  simplified) the strategy 
provided here is rather general and applies straightforwardly to other choices of
system parameters that produce the decoupling  of one or more normal modes. In fact
this method can extend this engineering of the  normal modes to more
complicated systems such as disordered harmonic lattices or complex networks. 

\section{Acknowledgements}

This work was funded by  the CSIC post-doctoral JAE program, the Balearic Government,
 the MICINN project FIS2007-60327 
and the MINECO project FIS2011-23526 and 	
FPI fellowship (BES-2012-054025). We acknowledge Gianluca Giorgi for discussions.

\appendix

\section{Analytical derivation of asymptotic entanglement}
\label{appA}

All the information about bipartite quantum correlations for a Gaussian continuous-variable state is condensed in its covariance matrix defined through the ten 
second order moments of $q_{(A, B)}$ and $p_{(A, B)}$ (in our case first order moments are initially zero). This bipartite covariance matrix 
defined for a system of two oscillators A and B, can be written as:
\begin{equation}\label{covariance_matrix}
V_{A B}= \left( {\begin{array}{cc}
 \alpha & \gamma  \\
 \gamma^{t} & \beta
\end{array} } \right)
\end{equation}
where $ \alpha,\beta,\gamma$ are $(2\times2)$ blocks: $ \alpha (\beta)$ contains the second order moments of oscillator subsystem A (B), 
and $ \gamma$ contains correlations of both subsystems. The minimum symplectic eigenvalue (of the  covariance matrix corresponding
to the partially transposed density matrix),  necessary to calculate the logarithmic
negativity, is given by:
\begin{equation}\label{nu}
\nu_{-} = \sqrt{\frac{1}{2}(a + b - 2g - \sqrt{(a + b- 2g)^2 - 4s)}}
\end{equation}
with $a= 4 \det(\alpha)$, $b= 4 \det(\beta)$ , $g= 4 \det(\gamma)$ and $s=16 \det V$. Normal modes coupled to the environment will reach in 
the asymptotic limit a thermal state, given by the Gibbs distribution $e^{-\frac{H_S}{T}}/\rm{Tr}( e^{-\frac{H_S}{T}})$. For a 
normal mode $(k)$ we get:
\begin{eqnarray} \label{asymptotic}
 \langle Q_k^2 \rangle_{th} = \frac{1}{2 \Omega_k} \coth \left(\frac{\Omega_k}{2 T}\right)  \nonumber \\
  \langle P_k^2 \rangle_{th} = \frac{\Omega_k}{2} \coth \left( \frac{\Omega_k}{2 T} \right)
\end{eqnarray}
being $\Omega_k$ the corresponding normal mode frequency and $T$ the reservoir temperature. On the other hand the uncoupled modes evolve 
freely. This means that the asymptotic covariance matrix can be calculated by expressing all second order moments of natural oscillators in 
terms of the normal modes and then substituting the asymptotic expressions corresponding to free modes or thermalized ones.

The covariance matrix in the asymptotic limit can be separated into three parts corresponding to the contributions of each normal mode. In 
terms of the blocks we can write: $\alpha= \sum_{i=1}^3 \mathcal{F}_{A i}^2 V_i$, $\beta= \sum_{i=1}^2 \mathcal{F}_{B i}^2 V_i$ and 
$\gamma = \gamma^T = \sum_{i=1}^2 {\mathcal{F}}_{A i} {\mathcal{F}}_{B i} V_i$. For a non-dissipative normal mode $(n)$ we have:
\begin{equation}\label{v_no_diss}
V_{no-diss}= \left( {\begin{array}{cc}
 \langle Q_n^2 \rangle & \frac{\langle \{ Q_n , P_n \} \rangle}{2}  \\
 \frac{\langle \{ Q_n , P_n \} \rangle}{2} & \langle P_n^2 \rangle 
\end{array} } \right)
\end{equation}
and for a dissipative one $(k)$ we get:
\begin{equation}\label{v_diss}
V_{diss}= \left( {\begin{array}{cc}
 \langle Q_k^2 \rangle_{th} & 0  \\
 0 & \langle P_k^2 \rangle_{th} 
\end{array} } \right)
\end{equation}
While elements in $V_{diss}$ are given by the expressions (\ref{asymptotic}) those of $V_{no-diss}$ are the ones corresponding to a free 
evolution of an harmonic oscillator. This analysis gives all the necessary elements in order to calculate asymptotic entanglement for pairs 
of oscillators in every particular situation in which one or two of the normal modes are un-coupled from the environmental action. 

Consider the specific case of an open chain $(\lambda_{1 3}=0)$ in which we have two equal frequencies $(\omega_1 = \omega_3 \equiv \omega)$ 
and two equal couplings $(\lambda_{1 2} = \lambda_{2 3} \equiv \lambda\neq 0)$ (Fig.
 \ref{figure1}a). In this case we get only one 
normal mode decoupled from the bath. In order to calculate the expression of the minimum symplectic eigenvalue we have to first calculate 
the elements of the three normal modes, that are shown here as vector columns:
\begin{equation} \label{modes_a}
 {\bf C}_{\delta} = \frac{1}{\sqrt{2}} \left( {\begin{array}{cc}
                            1\\
			     0  \\
			    -1
                           \end{array}} \right) 
;~
{\bf C}_{\pm} = c_{\pm} \left( {\begin{array}{cc}
                            \lambda \\
			    \Omega_{\pm}^2 - \omega^2  \\
			    \lambda
                           \end{array}} \right) \nonumber
\end{equation}
where we have labeled the non-dissipative mode as $\delta$ and the other two modes as
$\{\pm \}$. The corresponding normal modes frequencies  are $\Omega_\delta = \omega$
and $\Omega_{\pm} = \sqrt{(\omega_2^2 + \omega^2)/2 \pm \sqrt{\epsilon}}$, defining
$\epsilon = (\frac{\omega_2^2  - \omega^2}{2})^2 + 2\lambda^2$, and $c_{\pm}$ is
nothing but a normalization constant. We can now obtain all the terms appearing in
$V_{\pm}$.

The initial condition given in Eq. (\ref{initial_condition_original}) can be now
re-written in terms of the  non-dissipative normal mode as  $\langle Q_\delta^2 (0)
\rangle = \frac{1}{2 \omega} e^{-2 r}$ , $\langle P_\delta^2 (0) \rangle =
\frac{\omega}{2} e^{2 r}$ and $\langle  \{ Q_\delta, P_\delta \}(0) \rangle = 0$, and
then their free evolution is given by:
\begin{eqnarray}\label{free_evolution}
\langle Q_{\delta}^{2}\rangle &=& \frac{1}{2 \omega } (e^{2r}\sin^{2}(\omega t) + e^{-2r}\cos^{2}(\omega t))   \nonumber\\
\langle P_{\delta}^{2}\rangle &=& \frac{\omega}{2} (e^{-2r}\sin^{2}(\omega t) + e^{2r}\cos^{2}(\omega t))   \nonumber\\
\langle \{Q_\delta,P_\delta\} \rangle &=& 2 \sinh(2r) \cos(\omega t)\sin(\omega t)
\end{eqnarray}
where we have already used that $\Omega_\delta = \omega$. By substituting in
$V_{no-diss}$ (Eq. \ref{v_no_diss}) we can now obtain the  expressions of the
determinants and introducing into equation (\ref{nu}) yields the following expression
for the minimum symplectic  eigenvalue:
\begin{eqnarray} 
\label{nu_as}
 \frac{\nu_{-}(t)^2}{ 2 \lambda^2} &=& {\cal{G}}_0 + {\cal{G}}_1\cos(2 \omega t) - \nonumber \\ 
&-& \sqrt{\left( {\cal{G}}_0 + {\cal{G}}_1\cos(2 \omega t) \right)^2 - 4\sigma_P \sigma_Q } 
\end{eqnarray}
which is an oscillatory function with frequency $2\omega$ defined by the following
functions:
\begin{eqnarray}
 {\cal{G}}_0 = (\sigma_Q + \sigma_P) \cosh(2 r) \nonumber \\
 {\cal{G}}_1 = (\sigma_Q - \sigma_P) \sinh(2 r) \nonumber
\end{eqnarray}
and the dependence on bath temperature and shape of dissipative normal modes is given
by:
\begin{eqnarray}
\label{sigmas}
 \sigma_P &= \frac{\Omega_{+}}{2 \omega}  c_{+}^2 \coth\left( \frac{\Omega_{+}}{2 T} \right) + \frac{\Omega_{-}}{2 \omega}  c_{-}^2 \coth\left( \frac{\Omega_{-}}{2 T} \right) \nonumber \\
 \sigma_Q &= \frac{\omega}{2 \Omega_{+}}  c_{+}^2 \coth\left( \frac{\Omega_{+}}{2 T} \right) + \frac{\omega}{2 \Omega_{-}}  c_{-}^2 \coth\left( \frac{\Omega_{-}}{2 T} \right)
\end{eqnarray}
From Eq. (\ref{nu_as}), we can obtain the minimum entanglement (given when
$t=(2n+1)\frac{\pi}{2 \omega};~ n=1,2,3,...$) and the maximum one  (given when
$t=(n+1)\frac{\pi}{\omega};~ n=1,2,3,.. $) in order to recover Eq. (\ref{entan_1mode})
in section (\ref{asymptotic_ent}) of the main  text with the proper definitions
specified there.

On the other hand, if we move to situation represented
in Figure \ref{figure1}e by fixing $\lambda = \tilde{\lambda}_0$ 
(see Eq.  \ref{lambda0}) we have that the normal modes transform into:
\begin{equation} \label{modes_b}
 {\bf C}_{\delta} = \frac{1}{\sqrt{2}} \left( {\begin{array}{cc}
                            1\\
			    0  \\
			    -1
                           \end{array}} \right) 
;~
{\bf C}_{\epsilon} = \frac{1}{\sqrt{6}} \left( {\begin{array}{cc}
                            1\\
			    -2 \\
			     1
                           \end{array}} \right)
;~
{\bf C}_{CM} = \frac{1}{\sqrt{3}} \left( {\begin{array}{cc}
                           1 \\
			    1  \\
			    1
                           \end{array}} \right) \nonumber
\end{equation}
being the last one, ${\bf C}_{CM}$, the only dissipative mode. The corresponding normal frequencies are respectively: $\Omega_\delta = \omega$,
$\Omega_\epsilon = \sqrt{2\omega_2^2 - \omega^2}$ and $\Omega_{CM} = \sqrt{2 \omega^2 - \omega_2^2}$. Naturally we have to restrict ourselves 
to the regime $2\omega_3^2 > \omega > \omega_3^2/2$ in order for these quantities to be real and positive.

Keeping the same initial condition as in the previous case, we have that nothing changes in the expression of 
the free evolution of mode $\delta$ of Eq. (\ref{free_evolution}) while the free evolution of mode $\epsilon$ is given by:
\begin{eqnarray}
\langle Q_{\epsilon}^{2}\rangle &= \frac{2\omega_2 + \omega}{6 \Omega_\epsilon^2} e^{2r} \sin^{2}(\Omega_\epsilon t) + \frac{2\omega + \omega_2}{6 \omega \omega_2}e^{-2r}\cos^{2}(\Omega_\epsilon t)   \nonumber\\
\langle P_{\epsilon}^{2}\rangle &=\frac{2\omega + \omega_2 \Omega_\epsilon^2}{6 \omega \omega_2} e^{-2r} \sin^{2}(\Omega_\epsilon t) + \frac{2 \omega_2 + \omega}{6} e^{2r}\cos^{2}(\Omega_\epsilon t)   \nonumber\\
\langle \{Q_\epsilon,P_\epsilon\}\rangle &= \left( \frac{2\omega_2 + \omega }{3 \Omega_\epsilon}e^{2r} - \frac{(2\omega + \omega_2)\Omega_\epsilon}{3 \omega \omega_2} e^{-2r}\right) 
\cos(\Omega_\epsilon t)\sin(\Omega_\epsilon t) \nonumber
\end{eqnarray} 
We have assumed the same squeezing
parameter $r$ in the central oscillator of the chain (note that in the previous case the initial state of the central 
oscillator is not relevant and we have not specified it). Following the same procedure as above, we calculate an expression for the 
minimum symplectic eigenvalue. It is worth noting that in this case we have two contributions to the determinants of the free type 
$V_{no-diss}$ (Eq. \ref{v_no_diss}) corresponding to the two non dissipative modes, and only a dissipative one $V_{diss}$ (Eq. \ref{v_diss}) 
corresponding to the center of mass mode. 

The minimum symplectic eigenvalue yields:
\begin{eqnarray} 
\label{nu_two}
 2 \nu_{-}(t)^2 = & {\cal{A}}_0 + {\cal{A}}_1(t)- \\ \nonumber 
 & - \sqrt{\left( {\cal{A}}_0 + {\cal{A}}_1(t) \right)^2 - {\cal{B}}_0 - {\cal{B}}_1(t)} 
\end{eqnarray}
where we have defined the following quantities in order to simplify the expression. The constant terms:
\begin{eqnarray}
{\cal{A}}_0 &=& \cosh(2 r) \left( 4 (\sigma_Q + \sigma_P) + \mathcal{J}_{+}(\Omega_{\epsilon}^2 + \omega^2) \right) \nonumber \\
{\cal{B}}_0 &=& 64 \sigma_P \sigma_Q + \frac{4(\omega + \omega_2)^2}{81 \omega \omega_2} + \frac{32 \Omega_{\epsilon} \omega \mathcal{J}_{+}}{3} \left( 
\frac{\omega \sigma_P}{\Omega_\epsilon} + \frac{\Omega_\epsilon \sigma_Q}{\omega} \right) \nonumber
\end{eqnarray}
and the oscillating terms:
\begin{eqnarray}
 {\cal{A}}_1(t) & = & 4 \cos(2 \omega t) \sinh(2 r) (\sigma_Q - \sigma_P) + \nonumber \\
 &+& \mathcal{J}_{+} \cos(2 \omega t) \sinh(2 r)(\Omega_\epsilon^2 + \omega^2) + \nonumber \\
 &+&  \mathcal{J}_{-} \cos(2 \Omega_{\epsilon} t) \cosh(2r) (\Omega_\epsilon^2 - \omega^2) -  \nonumber \\ 
 &-& \mathcal{J}_{-} \cos(2 (\Omega_\epsilon - \omega) t) \sinh(2r) \frac{(\Omega_\epsilon + \omega)^2}{2} - \nonumber \\
 &-& \mathcal{J}_{-} \cos(2 (\Omega_\epsilon + \omega) t) \sinh(2r) \frac{(\Omega_\epsilon - \omega)^2}{2} \nonumber \\
  {\cal{B}}_1(t) &=& \cos(2 \Omega_\epsilon t) \frac{32 \Omega_{\epsilon} \omega \mathcal{J}_{+}}{3} \left( 
  \frac{\Omega_\epsilon \sigma_Q}{\omega} - \frac{\omega \sigma_P}{\Omega_\epsilon} \right) \nonumber
\end{eqnarray}
where different frequencies coming from the two non-dissipative modes are present. We have used $\mathcal{J}_{\pm} = \frac{1}{12 \omega} 
\left( e^{2 r} \frac{2\omega_2 + \omega}{\Omega_\epsilon^2} \pm e^{-2 r}\frac{2\omega + \omega_2}{\omega \omega_2} \right)$ and the 
two new bath dependent functions are now simply given by the contribution of the center of mass mode's thermal state:
\begin{eqnarray}
\label{sigmas2}
 \sigma_P = \frac{\Omega_{CM}}{6 \omega} \coth\left( \frac{\Omega_{CM}}{2 T} \right) \nonumber \\
 \sigma_Q = \frac{\omega}{6 \Omega_{CM}} \coth\left( \frac{\Omega_{CM}}{2 T} \right)
\end{eqnarray}

\section{Master Equation} 
\label{appB}

Assuming the general framework provided in section \ref{model}, we can postulate a
simple approach to the dynamical evolution in the  weak-coupling limit between
system and environment. By using the general Born and Markov approximations as well as
initial product state, a Markovian Master Equation (MME) for the reduced density
matrix of the open system can be obtained \cite{Breuer} in the normal modes  basis.
The resulting equation is not of the Lindbland form, thus complete positivity
(CP) is not guaranteed \cite{Rivas}. This issue  can be solved in two different ways,
either considering a rotating-wave-approximation (RWA) in the interaction Hamiltonian
(\ref{H_I})  $x_ix_j\rightarrow a_ia_j^\dagger+h.c.$ or performing a
strong-type RWA in the non-Lindbladian master equation by eliminating  oscillatory
terms of the form $e^{\pm i (\Omega_i \pm \Omega_j) t}$ that appear in the interaction
picture. The latter is the one we will  pursue. The advantages of this method 
not only
reside in obtaining a Master Equation in Lindbland form (thus CP), but also in that
dynamical  evolution can be solved analytically. However an exhaustive analysis in the
case of two harmonic oscillators shows a very well agreement  between results using
the original non-Lindbladian master equation and the strong RWA here used \cite{our}. 

The MME for the evolution of the reduced density matrix for a common bath in the strong RWA is then:
\begin{eqnarray} \label{MME} 
\frac{d \rho(t)}{dt}&=&-i[H_S,{\rho}(t)]- \nonumber\\
&-& \frac{i}{4} \sum_{n} \Gamma_{n} \left( [Q_n,\{P_n,\rho(t)\}] -[P_n,\{Q_n,\rho(t)\}]  \right) + \nonumber\\
 &+& D_{n}  \left( [Q_n,[Q_n,\rho(t)]] -  \frac{1}{\Omega_{n}^2} [P_n,[P_n,\rho(t)]] \right) 
\end{eqnarray}
Here $\Gamma_n$ and $D_n$ are constant coefficients (by virtue of the Markov approximation) accounting for the damping and diffusion effects 
respectively. Note that under this approximation, each normal mode is dissipating separately to the bath, i.e. they have independent decay 
rates. The bath has been considered to be in a thermal (Gibbs) state at temperature $T$, and to be composed by a continuum of 
frequencies characterized by an spectral density $J(\varOmega)$. For simplicity it has been considered to be Ohmic with a sharp cutoff 
$J(\varOmega) = \frac{2 \gamma}{\pi} \varOmega ~\Theta(\Lambda - \Omega)$, where $\Theta(x)$ is the Heaviside step function, $\Lambda$ is the 
largest frequency present in the environment (cutoff frequency) and $\gamma$ is a constant quantifying the strength of system-environment 
interaction (thus in the weak-coupling we have always that $\gamma$ $\ll$ $\Omega_i$ $\forall i \in \{1, ..., N\}$). This assumptions leads 
to the following definitions of the Master Equation coefficients:
 \begin{eqnarray} \label{Gammas}
&\Gamma_{n}&= \kappa_n^2 ~\frac{\pi}{2} \frac{J(\Omega_n)}{\Omega_n} = \gamma ~\kappa_n^2 \\ \label{Ds}
&D_{n}& = \kappa_n^2 ~\frac{\pi}{2} J(\Omega_n) \coth \left(\frac{\Omega_n}{2 T} \right)  = \gamma ~\kappa_n^2 \Omega_n \coth 
\left(\frac{\Omega_n}{2 T} \right) \nonumber
\end{eqnarray}
where we also assume $\Omega_i < \Lambda ~\forall i \in \{1,...,N\}$.

As we are interested in classical and quantum correlations of the system oscillators, a description for the evolution of the first and 
second order moments is necessary. This can be written in a simple form as:
\begin{equation}
 \dot{\bf{R}}= \mathcal{M}{\bf{R}}+{\bf N}. \nonumber
\end{equation}
where $\bf{R}$ is a column vector containing the $M= (2N + 1)N$ independent second order moments of the normal modes. The matrix 
$\mathcal{M}$ condenses all the information about their dynamical evolution and ${\bf N}$ determines the stationary values for long 
times (when $\dot{\bf{R}} = 0$). The dynamics of $\bf{R}$ can be solved in terms of the eigenvalues of $\mathcal{M}$:
\begin{equation}\label{mus}
 \{{\mu_{i j}}\}=\{ -\frac{\Gamma_{i}+\Gamma_{j}}{2} \pm i\left|\Omega_i \pm \Omega_j\right| \} ~;~i \leq j
\end{equation}
where the $i=j$ eigenvalues determine the evolution of $\langle Q_i^2 \rangle$, $\langle P_i^2 \rangle$ and $\langle \{Q_i,P_i\} \rangle$, 
while the ones with $i \neq j$ determine that of $\langle Q_i Q_j \rangle$, $\langle P_i P_j \rangle$ and $\langle \{Q_i,P_j\} \rangle$. Note 
that by virtue of eqs.(\ref{Gammas}) and (\ref{mus}) the decay of the normal modes is entirely governed by the effective couplings mentioned 
above, thus differences in their magnitude produce disparate temporal scales for the dissipation and diffusion of normal modes.

The stationary state is found to be $(\bf{R}_\infty = \mathcal{M}^{-1} {\bf N})$:
\begin{eqnarray} \label{asymptotic2}
 \langle Q_i^2 \rangle_\infty = \frac{D_i}{2 \Gamma_i \Omega_i^2}= \frac{1}{2 \Omega_i} \coth \left(\frac{\Omega_i}{2 T}\right)  \nonumber \\
  \langle P_i^2 \rangle_\infty = \frac{D_i}{2 \Gamma_i} = \frac{\Omega_i}{2} \coth \left( \frac{\Omega_i}{2 T} \right) \nonumber
\end{eqnarray}
being all the other second order moments equal to zero. Note that these expressions for the asymptotic limit recover the thermal state 
of the system at the bath temperature $T$ given by the Gibbs distribution in Eqs.(\ref{asymptotic}).

\end{document}